\documentstyle[aps,preprint,epsfig]{revtex}  
\tightenlines  

\newcommand{\be}{\begin{equation}} 
\newcommand{\ee}{\end{equation}} 
\newcommand{\bes}{\begin{eqnarray}} 
\newcommand{\ees}{\end{eqnarray}} 
\newcommand{\bfig}{\begin{figure}} 
\newcommand{\efig}{\end{figure}} 
\newcommand{\eps}{\epsilon} 
 
\newcommand{\et}{(\epsilon, \tau)}

\begin{document}       
  
\title{Exit-Times and {\Large $\epsilon$}-Entropy for Dynamical Systems,
Stochastic Processes, and Turbulence}
 
\author{M. Abel$^{(a)}$
 L. Biferale$^{(b)}$, M. Cencini $^{(c)}$, M. Falcioni$^{(a)}$,
  D. Vergni $^{(a)}$  and A. Vulpiani $^{(a)}$}

\address{$(a)$ Dipartimento di Fisica, Universit\'a di Roma ''La Sapienza''
 and INFM \\
p.le Aldo Moro 2, I-00185 Roma, Italy}
\address{$(b)
$ Dipartimento di Fisica, Universit\'a di Roma ''Tor Vergata''
 and INFM\\
via della Ricerca Scientifca 1, 00133 Roma, Italy}
\address{$(c)
$ Max-Planck-Institut f\"ur Physik komplexer Systeme, \\ 
N\"othnitzer Str. 38  , 01187 Dresden, Germany} 

\maketitle

\begin{abstract}
We present a comprehensive investigation of $\epsilon$-entropy, $h(\epsilon)$, in
dynamical systems, stochastic processes and turbulence.
Particular emphasis is devoted on a recently proposed approach to the
calculation of the $\epsilon$-entropy based on the exit-time
statistics. The advantages of this
method are demonstrated in examples of deterministic diffusive maps,
intermittent maps, stochastic self-affine and multi-affine signals and
experimental turbulent data.  Concerning turbulence, the multifractal
formalism applied to the exit time statistics allows us to predict
that $h(\epsilon)\sim \epsilon^{-3}$ for velocity time measurement.
This power law is independent of the presence of intermittency and has
been confirmed by the experimental data analysis.  Moreover, we show
that the $\epsilon$-entropy density of a $3$-dimensional velocity
field is affected by the correlations induced by the sweeping of large
scales.

PACS: 05.45.+b  \hfill\break

\end{abstract}

\section{Introduction}
\label{sec:1}

Many sciences, ranging from geophysics to economics, share the crucial
problem of extracting information about the underlying dynamics of a
system through the analysis of data time series \cite{Kantz97}.  In
these investigations, a central role is played by the evaluation of
the complexity degree of a string of data as a way to probe the
underlying dynamics \cite{Grass86,Politi97}. Since the pioneering
works of Shannon on information theory \cite{Shannon48}, entropy has
been proposed as the proper mathematical tool to quantitatively
address such a question.

Nowadays, entropy constitutes a key-concept to answer questions
ranging from the more conceptual aim to distinguish a pure stochastic
evolution from a chaotic deterministic one to the more applied goal of
quantifying the degree of predictability at varying the space-time
resolution \cite{ER85,GP_PRA,GW93,ABCPV96,cfkov00}.  The latter
question is evidently of primary importance, e.g., to set the proper
resolution of the data accumulation rate in experimental settings or
to efficiently compress data which have to be stored or transmitted.

The distinction between stochastic and deterministic chaotic evolution
can be formalised by introducing the Kolmogorov-Sinai ($KS$) entropy,
$h_{KS}$ \cite{Kolmo58}. Let us consider a time series $x_t$ (with
$t=1,\dots,T$) where, for simplicity, the time is discretised but
$x_t$ is a continuous variable. By defining a finite partition of the
phase-space, where each element of the partition has diameter smaller
than $\epsilon$, and by recording for each $t$ the symbol (letter)
identifying the cell $x_t$ belongs to, one can code the time series
into a sequence of symbols out of a finite alphabet. Then, from the
probabilities of words of length $m$ ($m$-words) one can compute the
$m$-block entropy. Finally, one measures the information-gain in going
from $m$-words to $(m+1)$-words: in the limit of infinitely long words
($m \rightarrow \infty$) and of arbitrary fine partition ($\epsilon
\rightarrow 0$) one obtains $h_{KS}$, that is an entropy per unit time
\cite{ER85}. The value of $h_{KS}$ characterises the process which has
generated the time series.  For example, in a continuous stochastic
evolution, which reveals more and more unpredictable outcomes at
increasing the resolution, the $KS$-entropy is infinite.  On the other
hand, a regular deterministic signal is characterised by a zero
$KS$-entropy, since it is completely predictable after a finite number
of observations, at any given resolution. Between these two limiting
cases, a finite positive value of $h_{KS}$ is the signature of a
deterministic chaotic dynamics.  The $KS$-entropy measures the growth
rate of unpredictability of the evolution, which coincides with the
rate of information acquisition necessary to unambiguously reconstruct
the signal. However, the distinction between chaotic and stochastic
dynamics can be troublesome in practical application (see
\cite{cfkov00} for a related discussion).

Indeed, only in simple, low dimensional, dynamical systems the
$h_{KS}$ evaluation can be properly carried out. As soon as one has to
cope with realistic systems, e.g. geophysical flows, the number of
degrees of freedom is so large that it inhibits any definite statement
based on the $KS$-entropy evaluation. Moreover, even if one were
able to compute the $KS$-entropy of those systems, many interesting
features can not be answered by only knowing $h_{KS}$.  As a
relevant example we mention the case of turbulence, the dynamics of
which is characterised by a hierarchy of fluctuations with different
characteristic times and spatial scales \cite{frisch}.  In this
respect the $KS$-entropy is related only to the fastest time scale
present in the dynamics.  Therefore, to quantify the predictability
degree depending on the range of scales and frequencies analysed, we
need a more general tool \cite{GW93,ABCPV96,GW_PRA}.

In order to make a step to overcome these difficulties, we consider a
scale-dependent quantity, namely the $\epsilon$-entropy,
$h(\epsilon)$, originally introduced by Shannon \cite{Shannon48} and
Kolmogorov \cite{Kolmo56} to characterise continuous processes. It is
remarkable that, in spite of its deep relevance for the
characterisation of stochastic processes and non trivial dynamical
systems, the $\epsilon$-entropy is not widely used in the physical
community.  Only recently, mainly after the review paper of Gaspard
and Wang \cite{GW93} and the introduction of the Finite Size Lyapunov
Exponent \cite{ABCPV96}, there appeared some attempts in the use of
the $\epsilon$-entropy.  For this reason, in Section \ref{sec:2} we
give a brief pedagogical review, aimed to introduce the reader to the
$\epsilon$-entropy and $(\epsilon,\tau)$-entropy. Practically the
$(\epsilon,\tau)$-entropy, $h(\epsilon,\tau)$, is the Shannon entropy
of time series sampled at frequency $\tau^{-1}$ and measured with an
accuracy $\epsilon$ in the phase space. 

We will see that the analysis of the $\epsilon$-dependence of
$h(\epsilon)$ is able to highlight many dynamical features of very
high-dimensional systems like turbulence as well as of stochastic
processes \cite{GW93,GW_PRA}.  The determination of $h(\epsilon,
\tau)$ is usually performed, as already stated, by looking at the
Shannon entropy of the coarse-grained dynamics on a $(\epsilon, \tau)$
grid in phase-space.  Unfortunately, this method suffers of so many
computational drawbacks that it is almost unusable in many interesting
situations. In particular, it is very inefficient when one
investigates phenomena arising from the complex interplay of many
different spatial and temporal scales, the ones we are interested in.
Therefore, here we resort to a recently proposed method \cite{ABCFVV00}
based on the {\it exit-time} analysis, which has been demonstrated to
be both practically and conceptually advantageous with respect to the
standard one.  In a few words, the idea consists in looking at a
sequence of data not at fixed sampling time but at fixed fluctuation,
i.e. when the signal is larger than some given threshold, $\epsilon$.
This procedure allows a noticeable improvement of the computational
possibility to measure the $\epsilon$-entropy. We give an ample
demonstration of the advantages of this method in a number of examples
ranging from one-dimensional dynamical systems, to stochastic (affine
and multi-affine) processes and turbulence.

As far as turbulence is concerned, we present both an application to
experimental data analysis and a theoretical remark.  Namely, we will
see that from the computation of the $\epsilon$-entropy of turbulent
flows one has a deep understanding of the spatial correlation induced
by the sweeping of large scales on the smaller ones. In order to
understand these features we also introduce and discuss a new
stochastic model of turbulent flows which takes into account
sweeping effects.

The paper is organised as follows. In Section \ref{sec:2} we briefly
define the $\epsilon$-entropy and discuss its properties; we use a
simple example which shows the conceptual relevance of this quantity
together with the difficulties of its computation.  In section
\ref{sec:3} we introduce the {\it exit-time} approach to the
calculation of the $\epsilon$-entropy discussing in detail its
theoretical and numerical advantages. In section \ref{sec:4}, we
discuss the use of the $\epsilon$-entropy in characterising
intermittent low dimensional dynamical systems and stochastic (affine
and multi-affine) processes.  In section \ref{sec:5} we present a
study of high-Reynolds experimental data and a theoretical analysis of
the $\epsilon$-entropy in turbulence.  Some conclusions and remarks
follow in section \ref{Sec:6}.  Details on the stochastic model
of a turbulent field are discussed in the Appendices.

\section{The {\Large $\epsilon$}-entropy}
\label{sec:2}
Assume a given time-continuous record of one observable, $x(t) \in
{\bf {\rm I}\!{\rm R}}$, over a total time $T$ long enough to ensure a
good statistics.  For the sake of simplicity, we start considering $x$ as
an observable of a 1d system.

The estimate of the entropy of the time record $x(t)$ 
requires the construction of a symbolic dynamics \cite{Shannon48,ER85,GW93}. 
With this purpose, one considers, as a first step, a grid
on the time axis, by introducing a small time-interval, $\tau$, so as
to obtain a sequence $\{x_i=x(t_i), \; i=1,\dots,N \}$,
with $N=[T/\tau]$ ($[\cdot]$ denotes the integer part). As a second
operation, one performs a coarse-graining of the phase space, with a
grid of mesh size $\eps$, and defines a set of symbols, $\lbrace S
\rbrace$ (the letters of the alphabet), that biunivocally correspond
to the so formed cells. Then, one has to consider the different words
of length $n$, out of the complete sequence of symbols:
$$
W^n_k(\epsilon, \tau)= \left( S_k,S_{k+1} , \dots, S_{k+n-1} \right),
$$
where $S_j$ labels the cell containing $x_j$. 
See Fig.~1 where the above codification is sketched.

From the probability distribution $ P(W^{n}(\epsilon,\tau))$, estimated
from the words frequencies, one calculates the block entropies
$H_n(\epsilon,\tau)$:
\begin{equation}
\label{eq:2-4}
H_{n} (\epsilon,\tau) = - \sum _{ \lbrace W^{n}(\epsilon,\tau )\rbrace }
P(W^{n}(\epsilon,\tau))  \ln P(W^{n}(\epsilon,\tau))\,,
\end{equation}
where $\lbrace W^{n}(\epsilon,\tau )\rbrace$ indicates the set of all
possible words of length $n$.  The $(\epsilon ,\tau)$-entropy per unit
time, $h(\epsilon , \tau)$, is finally defined as:
\begin{eqnarray}
\label{eq:2-3a}
h_n(\epsilon , \tau)&=& {1 \over \tau} \lbrack H_{n+1} (\epsilon,\tau)
-H_n (\epsilon,\tau) \rbrack \; ,\\
\label{eq:2-3b}
h(\epsilon , \tau) &=& \lim _{n \to \infty} h_n(\epsilon , \tau) =
{1 \over \tau} \lim _{n \to \infty} {1 \over n} H_{n} (\epsilon,\tau)\; .
\end{eqnarray}
For practical reasons the dependence on the details of the partition
is ignored, while the rigorous definition is given in terms of the
infimum over all possible partitions with elements of diameter smaller
than $\epsilon$ \cite{ER85,GW93}. Note that the above defined
$(\epsilon,\tau)$-entropy is nothing but the Shannon-entropy of the
sequence of symbols $\{ S_i\}$.  In the case of the time-continuous
evolutions, whose realizations are continuous functions of time, the
$\tau$ dependence of $h(\epsilon, \tau)$ does not exist
\cite{ER85,Bill}.  When this happens, one has a finite
$\epsilon$-entropy per unit time, $h(\epsilon)$. For genuine
time-discrete systems, one can simply put $h(\epsilon)\equiv
h(\epsilon,\tau=1)$. In all these cases
\begin{equation}
\label{eq:2-5}
h_{KS} = \lim_{\epsilon \to 0} h(\epsilon)\,,
\end{equation}

The determination of $h_{KS}$ involves the study of the limits $n\to
\infty$ and $\epsilon \to 0$ which are in principle independent, but
in all practical cases one has to find an optimal choice of the
parameters such that the estimated entropy is close to the exact value
\cite{Kantz97,cfkov00}.

For a genuine chaotic system, one has $0< h_{KS} < \infty $, i.e. the rate of
information creation is finite.  On the other hand, for a continuous
random process $h_{KS}=\infty$.  Therefore, in order to distinguish
between a purely deterministic system and a stochastic system it is
necessary to perform the limit $\epsilon \rightarrow 0$.
Unfortunately, from a physical or numerical point of view this is
extremely difficult.  Nevertheless, by looking at the behaviour of the
$\epsilon$-entropy of the signal at varying $\epsilon$ one can have
some qualitative and quantitative insights on the chaotic or
stochastic nature of the underlying process \cite{cfkov00}.
Moreover, for some stochastic processes one can explicitly give an
estimate of the entropy scaling behaviour of $\epsilon$-entropy
\cite{GW93}.  For instance, in the case of a stationary Gaussian
process with spectrum $S(\omega)\propto \omega^{-2}$, Kolmogorov
\cite{Kolmo56} has rigorously derived
\begin{equation}
\label{eq:2-6}
h(\epsilon) \sim {1 \over \epsilon^2} \; ,
\label{eq:kolmo56}
\end{equation}
for small $\epsilon$.
However, as we show in the following simple but non-trivial example there are many
practical difficulties in the computation of $h(\epsilon)$
\cite{cfkov00,ABCFVV00}.  Let us consider the chaotic map:
\begin{equation}
x_{t+1}=x_t + p\sin 2\pi x_t\,,
\label{eq:mappa}
\end{equation}
which for $p > 0.7326\dots$  produces a large scale
a diffusive behaviour \cite{diff}, i.e.:
\begin{equation}
\label{eq:3-4}
\langle \left( x_t -x_0 \right)^2 \rangle \simeq 2 \,\, D \,\, t 
\qquad {\mbox {for}} \qquad t \to \infty\; ,
\end{equation}
where $D$ is the diffusion coefficient.
By computing the $\eps$-entropy of this system one expects \cite{GW93,ABCFVV00}
\begin{equation}
\label{eq:3-2}
h(\epsilon) \simeq \lambda \,\,\,{\rm for} \,\,\, 
\epsilon \lesssim 1\;\;\;
h(\epsilon) \propto {D\over \epsilon ^2} \,\,\, {\rm for}
\,\,\, \epsilon \gtrsim 1 ,
\end{equation}
where $\lambda$ is the Lyapunov exponent.  In Fig.~2 we show that the
numerical computation of $h(\epsilon)$, using the standard
codification (Fig.~1) is highly non-trivial already in this simple
system. Indeed the behaviour (\ref{eq:3-2}) in the diffusive region is
just poorly obtained by considering the envelope of
$h_n(\epsilon,\tau)$ computed for different values of $\tau$; while
looking at any single (small) value of $\tau$ (one would like to put
$\tau=1$) one obtains a rather inconclusive result.  This is due to
the fact that one has to consider very large block lengths, $n$, in
order to obtain a good convergence for $H_{n+1}\et -H_{n}\et$ in
(\ref{eq:2-3b}).  In the diffusive regime, a dimensional
argument shows that the characteristic time of the system at scale
$\epsilon$ is $T_\epsilon \approx \epsilon^2 / D$.  If we consider for
example, $\epsilon = 10$ and typical values of the diffusion
coefficient $D \simeq 10^{-1}$, the characteristic time,
$T_{\epsilon}$, is much larger than the elementary sampling time
$\tau=1$.

Concluding this section, we remind that for systems living in
$d>1$ dimensions, the procedure sketched above, for the determination
of $h(\epsilon, \tau)$, goes unaltered, considering that the
set of symbols $\{ S \}$ now identifies cells in the $d$-dimensional
space where the state-vector ${\bf x}(t)$ evolves. 

\section{How to compute the {\Large $\epsilon$}-entropy with exit times}
\label{sec:3}

The approach we propose to calculate $h(\epsilon)$ differs from the usual
one in the procedure to construct the coding sequence of the signal at
a given level of accuracy \cite{ABCFVV00}.  
This is an important point because the quality of the coding
affects largely the result of the $\epsilon$-entropy computation.
An efficient procedure reduces redundancy and improves 
the quality of the results. The problem to encode
signals efficiently is quite old and widely discussed in the
literature \cite{Politi97,Welsh89}.
The most efficient compression or codification of
a symbolic sequence is linked to its Shannon entropy. The Shannon's
compression theorem \cite{Shannon48} states: given an alphabet with
$m$ symbols, and a sequence of these symbols, $\lbrace S_i,
i=1,\dots,N \rbrace$, with entropy $h$, it is not possible to
construct another sequence $\lbrace S_i^\prime, i=1,\dots,N^\prime
\rbrace$ -- using the same alphabet and containing the same infomation -- 
whose length $N^\prime$ is
smaller than $(h/\ln m)N$.  That is to say: $h/\ln m$ is the maximum
allowed compression rate.  As a consequence, if one is able to map a
sequence $\lbrace s_i, i=1,\dots,N_{\!s} \rbrace$ of $m$ symbols, into
another sequence $\lbrace \sigma_i, i=1,\dots,N_{\!\sigma} \rbrace$,
with the same symbols, the ratio $(N_{\!\sigma}/N_{\!s})\ln m$ gives
an upper bound for the entropy of $\lbrace s_i \rbrace$.  More
generally, if $\lbrace \sigma_i \rbrace$ is a codification of $\lbrace
s_i \rbrace$ without information loss, then the two sequences must
have equal total entropy: $N_{\!s} h(s) = N_{\!\sigma} h(\sigma)$.

Now we introduce the coding of the signal by the exit-time,
$t(\epsilon)$, that is the time for the signal to undergo a
fluctuation of size $\eps$.  To do so, we define an alternating grid
of cell size $\epsilon$ in the following way: we consider the original
continuous-time record $x(t)$ and a reference starting time $t=t_0$.
The first exit-time, $t_1$, is then defined as the first time necessary
to have an absolute variation equal to $\epsilon/2$ in $x(t)$, i.e.,
$|x(t_0 + t_1)-x(t_0)| \ge \epsilon/2 $. This is the time the signal
takes to exit the actual cell of size $\epsilon$.  Then we restart
from $t_1$ to look for the next exit-time $t_2$, i.e., the first time
such that $|x(t_0 + t_1 + t_2)-x(t_0 + t_1)| \ge \epsilon/2 $ and so
on, to obtain a sequence of exit-times: $\{t_i(\epsilon)\}$.  To
distinguish the direction of the exit (up or down out of a cell), we
introduce the label $k_i= \pm 1$, depending on whether the signal is
exiting above or below.  For clarifying the procedure see Fig.~3,
where we sketch the coding method for the signal shown in Fig.~1.

From Fig.~3 one recognises the alternating structure of the
grid: the starting point to find $t_{i+1}$ lies in the middle of the
cell $x(t_i)\pm \eps/2$, whereas it lies on the border of the cell
$x(t_{i-1})\pm \eps/2$.  In this way one avoids the fast exit out of a
cell due to small fluctuations (compare Figs.~1 and 3).  At the end of
this construction, the trajectory is coded without ambiguity, with the
required accuracy, by the sequence $\{ (t_i,k_i), \; i=1, \dots, M
\}$, where $M$ is the total number of exit-time events observed during
the total time $T$.  A continuous signal, evolving in a continuous
time, is now coded in two sequences -- a discrete-valued one $\{k_i\}$
and a continuous-valued one $\{t_i\}$. Performing a coarse-graining
of the possible values assumed by $t(\eps)$ by the resolution time $\tau_r$,
we accomplished the goal of obtaining a symbolic sequence.  After that, one
proceeds as usual, studying the ``exit-time words'' of various lengths $n$.
These are the subsequences of couples of symbols
\begin{equation}
\Omega^n_i (\epsilon, \tau_r)=\left(
    (\eta_i,k_i),(\eta_{i+1},k_{i+1}),
\dots , (\eta_{i+n-1},k_{i+n-1}) \right) \; ,
\end{equation}
where $\eta_j$ labels the cell (of width $\tau_r$) containing the
exit-time $t_j$.  From the probabilities of these words one calculates
the block entropies at the given time resolution,
$H^\Omega_n(\epsilon, \tau_r)$, and then the exit-time $(\epsilon,
\tau_r)$-entropies:
\begin{equation}
h^\Omega(\epsilon, \tau_r) = \lim_{n \to \infty}
H^\Omega_{n+1}(\epsilon, \tau_r) - H^\Omega_n(\epsilon, \tau_r)\; .
\end{equation}
The limit of infinite time-resolution gives us the $\epsilon$-entropy
{\it per exit}, i.e.:
\begin{equation}
h^\Omega(\epsilon) = \lim_{\tau_r \to 0}
h^\Omega(\epsilon, \tau_r)\,.
\label{homegalim}
\end{equation}
This result may be obtained also by arguing as follows.  There is a
one-to-one correspondence between the (exit-time)-histories and the
$(\epsilon,\tau)$-histories (in the limit $\tau \to 0$) originating
from a given $\epsilon$-cell. The Shannon-McMillan theorem
\cite{khinc} assures that the number of the typical
$(\epsilon,\tau)$-histories of length $N$, ${\cal N} (\epsilon,N)$, is
such that: $\ln {\cal N} (\epsilon,N) \simeq h(\epsilon) N \tau =
h(\epsilon) T$. For the number of typical (exit-time)-histories of
length $M$, ${\cal M} (\epsilon,M)$, we have: $\ln {\cal M}
(\epsilon,M) \simeq h^{\Omega}(\epsilon) M$. If we consider $T=M
\langle t(\epsilon) \rangle$ we must obtain the same number of (very
long) histories. Therefore, from the relation $M =
T/{\langle t(\epsilon) \rangle}$, where $\langle t(\eps)\rangle
=1/M\,\sum_{i=1}^M t_i$, we obtain finally for the $\epsilon$-entropy
per unit time:
\begin{equation}
h(\epsilon) = {M  h^\Omega(\epsilon) \over  T} =
\frac{h^\Omega(\epsilon)}{\langle t(\epsilon) \rangle}\,\,.
\label{epsent}
\end{equation}
Note that a relation similar to (\ref{epsent}), without the dependence
on $\epsilon$, has been previously proposed, in the particular case of
the stochastic resonance \cite{SR}. In such a case, where $x(t)$
effectively takes only the two values $\pm 1$ and the transition can
be assumed to be instantaneous, the meaning of the equation is rather
transparent.

At this point we have to remind that in almost all practical
situations there exists a minimum time interval, $\tau_s$, a signal
can be sampled with. Since there exists this minimum resolution time,
we can at best estimate $h^\Omega(\epsilon)$ by means of
$h^\Omega(\epsilon)= h^\Omega(\epsilon,\tau_s)$, instead of performing
the limit (\ref{homegalim}); so that we may put:
\begin{equation}
h(\epsilon) \simeq {h^\Omega(\epsilon,\tau_r) 
\over \langle t(\epsilon)\rangle }
\,\, ,
\label{bho}
\end{equation}
for small enough $\tau_r$.  In most of the cases, the leading
$\epsilon$-contribution to $h(\epsilon)$ in (\ref{bho}) is given
by the mean exit-time $\langle t(\epsilon) \rangle$ and not by
$h^\Omega(\epsilon,\tau_r)$. Anyhow, the computation of
$h^\Omega(\epsilon,\tau_r)$ is compulsory in order to recover, e.g., a
zero entropy for regular (e.g. periodic) signals.

Now we discuss how one can estimate the $\epsilon$-entropy in practice. 
In particular we introduce upper and lower bounds
for $h(\epsilon)$ which are very easy to compute in the exit time
scheme \cite{ABCFVV00}.  We use the following notation: for given
$\epsilon$ and $\tau_r$, $h^\Omega(\epsilon,\tau_r) \equiv h^\Omega(\{
\eta_i,k_i \})$, and we indicate with $h^\Omega(\{k_i\})$ and
$h^\Omega(\{ \eta_i\})$ respectively the Shannon entropy of the
sequence $\{k_i\}$ and $\{\eta_i\}$.  By applying standard results of
information theory \cite{Shannon48} one obtains:
\begin{enumerate}
\item[{\it a)}]
$
h^\Omega(\{ k_i \}) \leq h^\Omega(\{ \eta_i,k_i \})\,,
$ \hfill\break
since the mean uncertainty on the composed event $\{ \eta_i,k_i \}$
cannot be smaller than that of a partial one $\{ k_i \}$
(or $\{\eta_i\}$);
\item[{\it b)}]
$
h^\Omega(\{ \eta_i,k_i \})\leq h^\Omega(\{ \eta_i\}) + h^\Omega(\{k_i\}),
$ \hfill\break
since the uncertainty  is maximal if  $\{k_i\}$ and  $\{\eta_i\}$
are independent  (correlations can only decrease the uncertainty).
\end{enumerate}
Moreover, we observe that, for a given  finite resolution $\tau_r$, the
associated sequence $\{\eta_i\}$ satisfies the bound:
$$
h^\Omega(\{ \eta_i\}) \leq H^\Omega_1(\{ \eta_i\})\,\,,
$$
In the above relation $H^\Omega_1(\{\eta_i\})$ is the one-symbol
entropy of $\{ \eta_i\}$, (i.e. the entropy of the probability
distribution of the exit-times measured on the scale $\tau_r$) which
can be written as
$$
H^\Omega_1(\{ \eta_i\}) =  c(\epsilon) +
\ln \left( {\langle t(\epsilon)\rangle \over \tau_r} \right)\,\,,
$$
where $c(\epsilon) = -\int p(z)\ln p(z) {\mathrm d}z$, and $p(z)$ is
the probability distribution function of the rescaled exit-time
$z(\epsilon) = t(\epsilon)/\langle t(\epsilon)\rangle$.  Finally,
using the previous relations, one obtains the following bounds for the
$\epsilon$-entropy:
\begin{equation}
\label{bound-entro}
{h^\Omega(\{ k_i \}) \over \langle t(\epsilon) \rangle} \leq
h(\epsilon) \leq
{h^\Omega(\{ k_i \}) + c(\epsilon) + \ln
(\langle t(\epsilon)\rangle / \tau_r)
\over \langle t(\epsilon) \rangle} \,.
\end{equation}
Note that such bounds are relatively easy to compute and give a good
estimate of $h(\epsilon)$. The Equations (\ref{epsent}-\ref{bound-entro}) 
allow for a remarkable improvement of the computational efficiency.
Especially as far as the scaling behaviour of $h(\epsilon)$ is
concerned, one can see that the leading contribution is given by
$\langle t(\epsilon) \rangle$, and that $h^{\Omega}(\epsilon,\tau_r)$
introduces, at worst, a sub-leading logarithmic contribution
$h^\Omega(\epsilon,\tau_r) \sim \ln (\langle t(\epsilon) \rangle /\tau_r)$
(see eq.~(\ref{bound-entro})).  This fact is evident in the case of
Brownian motion.  In this case one has $\langle t(\epsilon) \rangle
\propto \epsilon ^2 / D$, and 
\begin{enumerate}
\item[(i)] 
$c(\epsilon)$ is $O(1)$ and independent of $\epsilon$
(since the Brownian motion is a self-affine process);
\item[(ii)] $h^\Omega(\{ k_i \}) \leq \ln 2$, is small compared with
$\ln(\langle t(\epsilon) \rangle /\tau_r)$. So that, neglecting the
logarithmic corrections, $h(\epsilon) \sim 1/\langle t(\epsilon)
\rangle \propto D \epsilon^{-2}$.
\end{enumerate}
In Fig.~4 we show the numerical evaluation of the bounds
(\ref{bound-entro}) for the diffusive map (\ref{eq:mappa}).  Fig.~4
has to be compared with Fig.~2, where the usual approach has been
used.  While in Fig.~2 the expected $\epsilon$-entropy scaling is just
poorly recovered as an envelope over many different $\tau$, within the
exit time method the predicted behaviour is easily recovered in all
the range of $\epsilon >1$ with a remarkable improvement in the
quality of the result.

We underline that the reason for which the exit time approach is more
efficient than the usual one is {\it a posteriori} intuitive.  Indeed,
at fixed $\epsilon$, $\langle t(\epsilon) \rangle$ automatically gives
the typical time at that scale, and, as a consequence, it is not
necessary to reach very large block sizes -- at least if $\epsilon$ is
not too small.  Especially for large $\epsilon$, we found that small
word lengths are enough to estimate the $\epsilon$-entropy accurately.
Of course, for small $\epsilon$ (i.e. the plateau of Fig.~4) one has
to use larger block sizes: here the exit time is $O(1)$ and one falls
back to the problems of the standard method.

For small $\epsilon$ in deterministic system one has to distinguish
two situations.
\begin{enumerate}
\item [{\it(a)}] $\epsilon \to 0$ for discrete-time systems. \hfill\break 
In this limit the exit-time approach coincides
with the usual one.  The exit-times always coincide with the minimum
sampling time, i.e.  $\langle t(\epsilon\to 0) \rangle \sim 1$ and we
have to consider the possibility to have jumps over more than one
cell, i.e., the $k_i$ symbols may take values $\pm 1,\pm 2, \dots$.
\item[{\it (b)}] $\epsilon \to 0$ for continuous-time systems. \hfill\break 
At very small $\epsilon$, due to the deterministic
character of the system, one has $\langle t(\epsilon)\rangle \sim
\epsilon$, and therefore one finds words composed with highly
correlated symbols. So one has to treat very large blocks in computing
the entropy \cite{Grass96}.
\end{enumerate}

However, as far as high dimensional systems are concerned, for some
aspects, the points (a) and (b) are not of practical interest.  
In these systems the analysis of the $\epsilon \to 0$ limit is usually
unattainable for several reasons \cite{GW93,cfkov00}, and, moreover,
in many cases one is more interested in the large $\eps$ scale
behaviour.  We believe that in these cases the approach presented
here, is practically unavoidable.

We conclude this section with two further remarks.  First, up to now we
considered a scalar signal as the output of a one-dimensional system.
This fact only entered in the two-valuedness of the $k$-variable. If
we are given a vectorial signal $ {\bf x }(t) $, describing the
evolution of a $d$-dimensional system, we have only to admit $2d$
values for the direction-of-exit variable $k$.  If the dynamics is
discrete one has also to consider the possibility of jumps over more
than one cell (see previous discussion).  

Second, one can wonder about the dependence of $h(\epsilon)$ on the
used observable.  Rigorous results insure that the Kolmogorov-Sinai
entropy, i.e.  the limit $\epsilon \to 0$ of $ h (\epsilon) $ is an
intrinsic quantity of the considered system, its value does not change
under a smooth change of variables.  In the case of $(\epsilon,
\tau)$-entropy, in principle there could be dependencies on the chosen
function. However, one can see that at least the scaling properties
should not strongly depend on the choice of the observable. If $A(x)$
is a smooth function of $x$, such that the following property holds:
\begin{equation}
\label{observable}
c_1 |\delta x| \leq | A(x+ \delta x) - A(x) | \leq c_2 |\delta x|,
\end{equation}
with $c_1$ and $c_2$ finite constants, then there exist two
constants $\alpha _1$ and $\alpha _2$ such that
\begin{equation}
\label{A-entro}
h_x(\epsilon/\alpha_1, \tau) \leq h_A(\epsilon, \tau)
\leq h_x(\epsilon/\alpha_2, \tau),
\end{equation}
where $h_A(\epsilon, \tau)$ and $h_x(\epsilon, \tau)$ are the
$(\epsilon, \tau) $-entropies computed using the observable $A$ and
$x$, respectively. This result implies that if $h(\epsilon, \tau)$
shows a power-law behaviour as a function of $\epsilon$, $h(\epsilon, \tau)
\sim \epsilon ^{-\beta}$, the same behaviour, with the same exponent
$\beta$, must be seen when using another, smooth, observable in
the determination of the $(\epsilon, \tau) $-entropy.

\section{Application of the {\Large $\epsilon$}-entropy 
to deterministic and stochastic processes}
\label{sec:4} 

\subsection{An intermittent deterministic mapping}
\label{ssec:41} 

We discuss the application of exit-time approach
to the computation of $\eps$-entropy in strongly intermittent
low-dimensional systems.

In presence of intermittency, the dynamics is characterised by very
long, almost quiescent (laminar) intervals separating short intervals
of very intense (bursting) activity (see Fig.~5).  Already at a
qualitative level, one realises that coding the trajectory shown in
Fig.~5 at fixed sampling times (Section \ref{sec:2}) is not very
efficient compared with the exit times method, where the information
on the very long quiescent periods is typically stored using only one
symbol.

To be more quantitative, let us consider the following one dimensional
intermittent map \cite{PM}:
\begin{equation}
x_{t+1}=(x_t+a x_t^{\,z})\,\, {\mbox {mod}}\;1\,,
\label{eq:41.1}
\end{equation}
with $z>1$ and $a>0$. The invariant density is characterised by a
power law singularity near $x=0$, which is a marginally stable fixed
point, i.e.  $\rho(x) \propto x^{1-z}$.  For $z \geq 2$, the density
is not normalisable, and an interesting dynamical regime, the
so-called {\it sporadic chaos}, appears \cite{Pnacs}. Namely, for
$z\geq 2$ the separation between two close trajectories behaves as:
\begin{equation}
|\delta x_n| \sim \delta x_0 \exp\left[c n^{\nu_0} (\ln n)^{\nu_1}\right]\,,
\label{eq:interm}
\end{equation}
with $0<\nu_0<1$ or $\nu_0=1$ and $\nu_1<0$.  In the sporadic chaos regime,
nearby trajectories diverge with a stretched exponential, even if the
Lyapunov exponent is zero. For $z<2$ the system follows  the usual chaotic 
motion with $\nu_0=1$ and $\nu_1=0$.

Sporadic chaos is intermediate between chaotic motion and regular one. 
This can be understood by computing the Kolmogorov-Chaitin-Solomonoff 
complexity \cite{Pnacs}, or, as we show in the following, by studying 
the mean exit time.

By neglecting the contribution of $h^{\Omega}(\epsilon)$, and
considering only the mean exit time, we can estimate the total
entropy, $H_N$, of a trajectory of length $N$ as
\begin{equation}
H_N \propto {N \over \langle t(\epsilon)\rangle_N}  
\qquad {\mbox {for large}}\; N 	\,, 
\label{eq:compress}
 \end{equation} where $\langle [...]\rangle _N$ indicates that the
mean exit time is computed on a sequence of length $N$. Due to the
power law singularity at $x=0$, $\langle t(\eps) \rangle_N$ depends on
$N$.  In equation~(\ref{eq:compress}), we have dropped from $H_N$ the
dependence on $\epsilon$, which is expected to be weak.  Indeed, due
to singularity near the origin, one has that the exit times at scale
$\epsilon$ are dominated by the first exit from a region of size
$\epsilon$ around the origin. So that, $\langle t(\eps) \rangle_N$
approximately gives the duration of the laminar period (this is exact
for $\epsilon$ large enough).

In Fig.~6, the behaviour of  $\langle t(\epsilon) \rangle_N$  is shown
as a function of  $N$ and $z$ for two different choices of  $\epsilon$. 
For large enough $N$ the behaviour is almost 
independent of $\epsilon$, and for $z\geq 2$ one has
\begin{equation}
\langle t(\epsilon) \rangle_N \propto N^{\alpha}\,,\qquad{\mbox{where}}\qquad
\alpha={z-2 \over z-1}\,.
\label{eq:tave} 
\end{equation}
The value of $\alpha$ is obtained by the following argument:
the power law singularity leads to $x_t \approx 0$ most of the time, and
moreover, near the origin the map (\ref{eq:interm}) can be approximated by the 
differential equation $dx/dt=a x^z$ \cite{PM}.
Therefore, denoting with  $x_0$ the initial condition, one solves the
differential equation obtaining
$$
 (x_0+\epsilon)^{1-z}-x_0^{\,\,1-z}=a(1-z) t(\epsilon)\,.
$$
Now, due to the singularity, $x_0$ is typically much smaller than 
$x_0+\epsilon$, and 
hence we can neglect the term $(x_0+\epsilon)^{1-z}$,  so that 
the exit time is $t(\epsilon)\propto x_0^{1-z}$ .
By the probability density of $x_0$, $\rho(x_0)\propto
x_0^{1-z}$, one obtains the probability distribution of the exit times
$\rho(t)\sim t^{1/(1-z)-1}$, the factor $t^{-1}$ takes into account
the non-uniform sampling of the exit time statistics (see discussion
after equation (\ref{legendre2}).
Finally 
the average exit time on a trajectory of length $N$, which is given by
\begin{equation}
\langle t(\epsilon)\rangle _N \sim \int_0^N t\,\rho(t)\,{\rm d}t
\sim N^{z-2 \over z-1}\; .
\end{equation}

The total entropy is finally given by
$$
H_N\sim {N \over N^{z-2\over z-1}} \sim N^{1\over z-1}\,,
$$
note that this is exactly the same $N$-dependence found with the
computation of the 
algorithmic complexity \cite{Pnacs}. Let us underline that the
entropy per unit time goes to zero very slowly, because of the sporadicity
$$
{H_N \over N} \sim {1 \over \langle t(\epsilon)\rangle_N}\,.
$$ 
Let us note that we arrive at this results without any partitions
of the phase space of the system.
\subsection{Affine and multi-affine stochastic processes}

Self-affine and multi-affine processes are fully characterised by the 
scaling laws of the moments of signal increments \cite{frisch,PV87,BJPV98},
$\delta_{t}x = x(t_0) - x(t_0+ t)$~:
\begin{equation}
   \langle \langle |\delta_{t}x(t_0)|^q \rangle \rangle \sim t^{\,\zeta(q)} \,\,,$$
   \label{scalaff}
\end{equation}
where $\zeta(q)$ is a linear function of $q$, $\zeta(q) = \xi \,q $, for a self-affine signal 
($\xi$ is the H\"older exponent characterising the process)
and a non-linear function of $q$ for a multi-affine signal. 
The average $\langle \langle \cdot\rangle \rangle $ is defined as
the average over the process distribution
 $P(t_0, \delta_t x(t_0))$, which gives the probability to have a 
fluctuation, $\delta_t x(t_0)$, at the instant $t_0$. In the case of a 
stationary process, as it will be always assumed here, the probability
distribution is time 
invariant and the average  $\langle \langle \cdot\rangle \rangle $
is computed by invoking an ergodic hypothesis, as a time-average. 

Sometimes, with an abuse of language, a multi-affine process is also
called a multi-fractal process. 
While a self-affine process has a global scaling-invariant 
probability distribution function, a  multi-affine process 
can be constructed by  requiring a local (in time) scaling 
invariant fluctuations \cite{frisch}. 
In a nutshell, one assumes a spectrum of different local 
scaling exponents $\xi$: $\delta_{t} x(t_0) \sim t^{\,\xi(t_0)}$ 
with the probability $P_t(\xi) \sim t^{\,1-D(\xi)}$  
to observe a given H\"older exponent $\xi$ at time increment $t$.
The function $D(\xi)$
can  be interpreted as the fractal dimension of the set where the 
H\"older exponent $\xi$ is observed \cite{PV87}. 
The scaling exponents $\zeta(q)$ are related to $D(\xi)$ by a Legendre
transform. Indeed, one may define the average process as an
average over all possible singularities, $\xi$, weighted by the probability
to observe them: 
$$\langle\langle (\delta_t x)^q \rangle\rangle \sim 
   \int {\mathrm d} \xi\, t^{\xi q}\, t^{1-D(\xi)}\,\,,$$
which in the limit of small $t$ by a saddle point estimation becomes:
\begin{equation}
 \langle\langle(\delta_t x)^q\rangle\rangle \sim t^{\zeta(q)} \qquad {\mbox{with}} \qquad
   \zeta(q) = \min_\xi(q\xi + 1 - D(\xi))\,.
\label{legendre}
\end{equation}
Eq.~(\ref{legendre}) can be generalised to 
$\zeta(q) = \min_\xi(q\xi + d - D(\xi))$ if the considered signal 
is embedded in a d-dimensional space. \\
Let us notice that in this language, the already discussed Brownian motion 
corresponds to a self-affine signal with only one possible exponent $\xi=1/2$ 
with $D(1/2)=1$.
In Appendix A one finds how to construct 
arbitrary self-affine and multi-affine stochastic processes. 

Let us now investigate the $\epsilon$-entropy properties of these two
important classes of stochastic signals by using the exit-time
approach. We will proceed by discussing the general case of
multi-affine processes, noting that the self-affine one is a
particular case of them corresponding to have only one exponent in
the spectrum.

The exit-time probability distribution function
can be guessed by ``inverting'' the multifractal 
probability distribution functions \cite{BCVV99}.  
We expect that the following dimensional inversion should be correct 
(at least as far as leading scaling properties are  concerned). 
We argue that the probability to observe an exit of 
the signal through a barrier of height $\delta x$ in a time $t(\delta x)$
is given by 
$P_{\delta x} (t(\delta x)) \sim (\delta x)^{(1-D(\xi))/\xi}$
where the height of the barrier and the exit-time are related by
the inversion of the previously introduced multi-affine scaling relation
$t(\delta x) \sim (\delta x) ^{1/\xi}$. In this framework
we may write down the ``multifractal'' estimate \cite{BCVV99} of the
exit-time moments, also called inverse structure functions \cite{mogens}:
\begin{equation} \Sigma_q(\delta x) \equiv 
 \langle \langle t^q(\delta x )\rangle \rangle\ 
\sim \int {\mathrm d}\xi \, (\delta x)^{{q+1-D(\xi)}\over \xi} 
                      \sim (\delta x)^{\chi(q)}\;,
\label{multi_inv} 
\end{equation}
where $\chi(q)$ is obtained with a saddle point estimate in the limit of
small $\delta x$:
\begin{equation}
   \chi(q) = \min_\xi \left({{q + 1 - D(\xi)} \over \xi} \right)\;.
   \label{legendre2}
\end{equation}
The averaging by counting the number
of exit-time events $M$ (as we did in the previous sections) 
and the averaging  with the uniform ``multi-fractal'' distribution 
are connected by the following relation \cite{BCVV99}:
$$ 
\langle \langle t^q(\delta x)\rangle \rangle =\lim_{M \rightarrow \infty} 
   \sum_{i=1}^{M} t_i^q {t_i \over {\sum_{j=1}^{M} t_j}} = 
   {{\langle t^{q+1}(\delta x)\rangle }
 \over {\langle t(\delta x)\rangle}}\,\,.
$$
where the term $t_i / {\sum_{j=1}^{M} t_j}$ takes into account
the non-uniformity of the exit-time statistics.
From the previous relation evaluated for $q=-1$ we can easily  deduce 
the estimate for the mean exit-time scaling law:
\begin{equation}
\langle t(\delta x) \rangle= \langle \langle t^{-1}(\delta x) 
\rangle \rangle^{-1} \sim (\delta x)^{-\chi(-1)}
\label{pred1}
\end{equation}
and therefore, as in the previous sections, we may estimate 
the leading contribution to the  $\epsilon$-entropy of a multi-affine signal:
\begin{equation}
h(\delta x) \sim (\delta x)^{\chi(-1)}\,\,.
\label{pred2}
\end{equation}
Let us notice that in the simpler case of a self-affine signal
with H\"older exponent $\xi$, 
this is nothing but the dimensional estimate $h(\delta x) \sim (\delta
x)^{-1/\xi}$ which is rigorous for Gaussian processes \cite{Kolmo56}. 
In this case the above argument is also in agreement with the bounds
(\ref{bound-entro}): indeed for an affine signal the function $c(\eps)$
entering in (\ref{bound-entro}) does not depend on $\eps$
(we note here that $\delta x$ plays the same role of $\epsilon$).

In Fig.~7a-b we show the numerical estimate of the
bounds (\ref{bound-entro}) on the $\epsilon$-entropy
in two different self-affine signals with H\"older exponents $\xi=1/3$
and $\xi=1/4$ respectively 
(for details on the processes generation  see Appendix A). 
The agreement with the expected result is very good. 
Let us notice that with the usual approach 
to the calculation of the $\epsilon$-entropy for these simple
signals the detection of the scaling behaviour is not so easy 
(see Figures 15,16 and 17 of \cite{GW93}). 

In Fig.~8 we show the numerically computed lower and upper bounds 
for the $\epsilon$-entropy of a multi-affine signal by using
the mean exit-time estimate. The multi-affine signal here studied is 
characterised by having $\zeta(q)$ as the scaling exponent measured in 
turbulence (see next section). In particular, this means that $\zeta(3)=1$, and
using Eqs.~(\ref{legendre}) and (\ref{legendre2}) $\chi(-1)=-3$ independently on the
shape of $D(\xi)$. This is the $\eps$-entropies counterpart of
the Kolmogorov $4/5$ law~\cite{frisch}.

The agreement with the multifractal
prediction (the straight lines in Fig.~8) is impressive. 
To our knowledge this is the first direct estimate of $\epsilon$-entropy
in multi-affine signals. We stress that the non trivial aspect of such an 
estimate is contained in the derivation of the inverse multifractal
formulas (\ref{multi_inv})-(\ref{legendre2}).

\section{{\Large $\epsilon$}-entropy and exit times in turbulence} 
\label{sec:5} 
A turbulent flow is characterised by the presence of highly non-trivial
chaotic fluctuations in space and time \cite{frisch}.
The question we want to address here is to understand 
which kind of information can be captured by studying the $\epsilon$-entropy 
of this important high dimensional dynamical system.  
The main physical mechanism is the energy transfer from large scales, $L_0$,
i.e. scales where forcing is active, down to the dissipation scale, $\eta$,
where kinetic energy is converted into heat \cite{frisch,BJPV98}. 
The ratio between these two scales increases with the Reynolds number.  
Fully developed turbulence  corresponds to
the limit of very high Reynolds numbers.
In this limit, a turbulent velocity field develops scaling laws in 
the range of scale intermediate between $L_0$ and $\eta$, 
the so-called inertial range. 
Kolmogorov (1941) theory assumes a perfect 
self-similar behaviour for the velocity field in the inertial range. 
In other words, the velocity field was thought to be a continuous 
self-affine field with H\"older exponent $\xi=1/3$ as a function 
of its spatial coordinates: 
$$|v(x+R,t)-v(x,t)| \sim R^{1/3}\,\,, $$
(hereafter, for simplicity, we neglect the vectorial notation).
In terms of an averaged observable, this implies that the structure
functions, i.e. the moments of simultaneous velocity differences at distance $R$,
have a pure power-law dependency for $ \eta \ll R \ll L_0$:
\begin{equation}
S_p(R) = \langle\langle|v(x+R,t)-v(x,t)|^p\rangle\rangle \sim R^{\,\zeta(p)}\,.
\label{eq:sf}
\end{equation}
with $\zeta(p)=p/3$. Experiments and numerical simulations have
indeed shown that there are small (but important) corrections
to the Kolmogorov (1941) prediction. This problem
goes under the name of intermittency, the origin of which is
still one of the main open problem in the theory of Navier-Stokes
equations \cite{frisch,BJPV98,MY75}.
In the language of the previous section, an intermittent field is a 
multi-affine process. 

As far as the time-dependency of a turbulent velocity field is concerned, 
one can distinguish between two different time measurements.
First, the standard one (actually used in most of the experimental 
investigation), consists in measuring the time evolution by a probe 
fixed in some spatial location, say $x_p$, in the flow. The
time evolution obtained in this way is strongly affected by 
the spatial correlations induced by the large scales sweeping. As a result,
one can apply the so-called frozen-turbulent hypothesis (Taylor
hypothesis) \cite{MY75},
which connects a time-measurement with a spatial measurement by the following
relation:
$$ v(x_p,t_0+ t)-v(x_p,t_0) \sim v(x_p-R,t_0)-v(x_p,t_0) $$
where $R = t U_0$, where $U_0$ is the mean large scale 
sweeping velocity characteristic of the experiment. \\
As a result of the Taylor-hypothesis, one has that time-measurements
also show power-law behaviour with the same characteristic exponents
of the spatial measurements, namely, within the Kolmogorov theory:
$$ \langle \langle |v(x_p,t_0+ t)-v(x_p,t_0)|^p \rangle \rangle \sim t^{\,\zeta(p)}\,\,.$$
A second interesting possibility to perform time measurements 
consists in the so-called Lagrangian measurements \cite{lag2}.
In this case, one has to follow the trajectory of a single 
fluid particle and measuring the time properties 
locally in the co-moving reference frame. 
The main characteristics of this method is that the sweeping is removed 
and so one can probe in details the "proper" time-fluctuations 
induced by the non-linear terms of the Navier-Stokes equations 
(for recent theoretical and numerical
investigations of similar issues see \cite{lag2,lag1,bbct99}).\\
The phenomenological understanding of all these spatial and temporal
properties are well summarised by the Richardson-cascade. The cascade
picture describes a turbulent flow in terms of a superposition 
of fluctuations (eddies)
hierarchically organised on a set of scales ranging from the
largest one, $L_0$, to the smallest one, $\eta$,
say $\ell_n = 2^{-n}L_0$, with  
$n=0, \ldots, N_{max}$ and $N_{max}= \log_2(L_0/\eta)$. 
Each scale  has its own typical evolution time, $\tau_n$, 
given in terms of the velocity difference at
that scale, $\delta_nv = v(x+\ell_n)-v(x)$, by
the dimensional estimate: $\tau_n =  \ell_n/\delta_nv \sim (\ell_n)^{2/3}$.
The most relevant dynamical interactions are supposed to happen only between 
eddies of similar size, while each eddy is also subject
to the spatial sweeping from eddies at larger scales. 
The energy is transferred down-scale from the largest-eddy (the
mother) to its daughters and from the daughters to their
grand-daughters and so on in a multi-step process similar,
quantitatively and qualitatively to a stochastic multiplicative
process \cite{bbt98,bbcct99}.  \\

As a result of the previous picture, one can mimic a turbulent flow
with a stochastic process hierarchically organised in space, and with
suitable time-dependence able to reproduce both the overall sweeping
and the eddy-turn-over times hierarchy \cite{bbcpvv93,bbccv98}. In the
Appendices A and B we briefly remind a possible choice for these
stochastic process.

\subsection{Experimental data analysis}
\label{ssec:51}

We present now the computation of the $\epsilon$-entropy for two sets
of high Reynolds number experimental data, obtained from an experiment
in Lyon (at $Re_{\lambda}=400$) and from another experiment in Modane
(at $Re_{\lambda}=2000$).  The measurement in Lyon has been taken in a
wind tunnel with a working section of 3.0 m and a cross section of
(0.5 m)x(0.5 m).  Turbulence was generated by a cylinder placed inside
the wind tunnel, its diameter was 0.1 m. The hot wire was placed 2.0 m
behind the cylinder. The separation between both probes was
approximately 1 mm \cite{lyon}. The measurement in Modane has been
taken in a wind tunnel where the integral scale was $L \sim$ 20 m and
the dissipative scale was $r_{diss} = $ 0.3 mm.

Let us first make an important remark. Whenever one wants to apply the
multifractal formalism to turbulence there exist some analytical and
phenomenological constraints on the shape of the
function $D(\xi)$ entering in the multifractal description. 
In particular, the most important constraint is the exact result
$\zeta(3)=1$.  This, in turn, implies that independently of the
possible multifractal spectrum of the turbulent field one has
$\chi(-1)=-3$.  So that as stated in the previous section, one obtains:
\begin{equation}
   h(\epsilon)\sim \epsilon^{\,\chi(-1)}=\epsilon^{-3}\,\,,
\label{predturb}
\end{equation}
this is the $\epsilon$-entropy equivalent of the $\zeta(3)=1$ result,
i.e. of the $4/5$ law of turbulence \cite{frisch} (see equations
(\ref{pred1}) and (\ref{pred2})). This means that there are not
intermittent corrections to the $\epsilon$-entropy.  We have tested
this prediction (here for the first time presented), which has been already
confirmed in the analysis of the stochastic multi-affine signal in
section IV-B, in two different experimental data sets. \\
In Fig.~9 we show the $\epsilon$-entropy
computed for two different sets of experimental data. 
As one can see, the theoretical prediction $h(\epsilon) \sim \epsilon^{-3}$ is well 
reproduced only for large $\epsilon$ values, while for intermediate values 
the entropy shows a continuous bending without any clear scaling behaviour,
only when  $\epsilon$ reaches values corresponding to dissipative velocity
fluctuations we have the dissipative scaling
$\langle t(\epsilon) \rangle \sim \epsilon$. \\
The strong intermediate regime between the dissipative and the inertial
scaling behaviours is not a simple out-of-control finite Reynolds-effect.
In fact, within the multifractal model of turbulence, one can 
understand the large crossover between the two power laws
in terms of the so-called Intermediate-Dissipative-Range (IDR).  
The existence of an IDR was originally  predicted in \cite{fv} 
and furtherly analysed in \cite{BCVV99,pv,gc}.\\
The IDR brings the signature of the mechanism stopping the 
turbulent energy cascade, i.e. how viscous mechanism are effective
in dissipating turbulent energy. In particular, it was shown that 
the IDR can be fully described within the multifractal description
once one allows the possibility to have different viscous cut-off
depending on the local degree of velocity singularity, i.e. 
depending on the local realization of the $\xi$ scaling exponent. 
The main idea consists in using again the multifractal superposition
(\ref{multi_inv}) but considering that for velocity fluctuations
at the edge between the inertial and the viscous range not all
possible scaling exponents contribute to the average \cite{fv,BCVV99}.
It turns out that in the case of exit-time moments, the extension of the
IDR is much more important then what was previously measured for the
velocity structure functions (\ref{eq:sf}). 
Therefore, the strong finite-range effects showed
by the experimental data analysis of Fig.~9  can be qualitatively
and quantitatively  understood as an effect of the IDR \cite{BCVV99}.

Let us conclude this section by comparing our results with a previous 
study of the $\epsilon$-entropy in turbulence \cite{GW_PRA}. 
There it was argued the following scaling behaviour:
\begin{equation}
h(\epsilon) \sim \epsilon^{-2}\,,
\label{wrong}
\end{equation}
which differs from our prediction. The behaviour (\ref{wrong}) has been
obtained assuming that $h(\epsilon)$ at
scale $\epsilon$ is proportional to the inverse of the typical eddy
turnover time at that scale.  We remind that here $\epsilon$
represents a velocity fluctuation $\delta v$.  Since
the typical eddy turnover time for velocity fluctuations of order
$\delta v \sim \epsilon$ is $\tau(\epsilon) \sim \epsilon^2$,
the Eq.~(\ref{wrong}) follows. Recalling the discussion of section V-A
about the two possible way of measuring a turbulent time signal it is
clear that the scaling (\ref{wrong}) holds only in a Lagrangian
reference frame (see also \cite{ABCPV96}).  This explains
the difference of our prediction and (\ref{wrong}).

\subsection{An $\epsilon$-entropy analysis of the Taylor hypothesis in
fully developed turbulence}
\label{sec:52}

By studying the $\epsilon$-entropy for the velocity field of turbulent
flows in $3+1$ dimension, $h^{st}(\epsilon)$ ($st$ indicates {\it space} 
and {\it time}), we argue that the usually accepted Taylor hypothesis
implies a spatial correlation which can be quantitatively characterised 
by an ``entropy'' dimension ${\cal D}=8/3$. In this section, for the
sake of simplicity, we neglect intermittency, i.e. we assume
a pure self-affine field with H\"older exponent $\xi=1/3$.

We discuss now how to construct a
multi-affine field with the proper spatial and temporal scaling.
The idea consists in defining the signal as a dyadic three-dimensional 
superposition of  wavelet-like functions 
$\varphi(({\bf x} - {\bf x}_{n,{\bf k}}(t))/\ell_n)$ 
whose centres move according to a swept dynamics. 
The coefficients of the decomposition  $a_{n,{\bf k}}(t)$
are stochastic functions chosen with suitable self-affine scaling 
properties both in time and in space.
In particular, the exact definition for a field with
 spatial H\"older exponent $\xi$ in $d$ dimensions is (see Appendix A and B for
details):
\begin{equation}\label{aff-proc}
v({\bf x},t)= \sum_{n=1}^M \sum_{k=1}^{2^{d(n-1)}} a_{n,k}(t)\,
\varphi \! \left(\frac{{\bf x} - {\bf x}_{n,k}(t)}{\ell_n}\right)\,\,,
\end{equation}
where ${\bf x}_{n,k}$ is the centre of the $k^{th}$ wavelets at the level $n$
(for each dimension we consider one branching (i.e. two variables)
 for passing to the $n+1$ level, see Fig.~10).
According to the Richardson-Kolmogorov cascade picture, one assumes
that sweeping is present, i.e.,
${\bf x}_{n+1,k}={\bf x}_{n,k^\prime}+{\bf r}_{n+1,k}$ where
$(n,k^\prime)$ labels the ``mother'' of the
$(n+1,k)$-eddy and ${\bf r}_{n+1,k}$
is a stochastic vector which depends
on ${\bf r}_{n,k^\prime}$  and evolves with characteristic time
$\tau_n \propto (\ell_n)^{1-\xi} $.

If the coefficients $\{a_{n,k}\}$ and $\{ {\bf r}_{n,k}\}$
have characteristic time $\tau_n \sim (\ell_n)^{1-\xi}$ and $\{a_{n,k}\} \sim (\ell_n)^\xi$, 
it is possible to show (see Appendix A and B for details) that the field  
(\ref{aff-proc}) has the properties
\begin{eqnarray}
|v({\bf x}+{\bf R},t_0) - v({\bf x},t_0)|
&\sim& |{\bf R}|^{\xi}\;, \label{scaling-x}\\
|v({\bf x},t_0+t) - v({\bf x},t_0)|
&\sim& t^{\,\xi}\,\,; 
\label{scaling-t}
\end{eqnarray}
in addition the proper Lagrangian sweeping is satisfied.
Now we are ready for the $\epsilon$-entropy analysis of the field (\ref{aff-proc}).
If one wants to look at the field $v$ with a resolution $\epsilon$,
one has to take $n$ up to $N$ given by:
\begin{equation}
(\ell_N)^{\xi} \sim \epsilon\;,
\end{equation}
in this way we are sure to consider velocity fluctuations of order $\epsilon$.
Then the number of terms contributing to (\ref{aff-proc}) is
\begin{equation}
\#(\epsilon) \sim (2^d)^N \sim \epsilon^{-d/\xi}\;.
\end{equation}
By using a result of Shannon \cite{Shannon48} one estimates
the $\epsilon$-entropy of the process $a_{n,k}(t)$ (and
also of ${\bf r}_{n,j}$) as:
\begin{equation}
h_n(\epsilon) \sim \frac{1}{\tau_n} \log \left ({1 \over \epsilon} \right)\;,
\end{equation}
where the above relation is rigorous if the processes $a_{n,k}(t)$
are Gaussian and with a power spectrum different form zero on
a band of frequency $\sim 1/\tau_n$.
The terms which give the main contribution are those with $n \sim N$ with
$\tau_N \sim (\ell_N)^{1-\xi}  \sim \epsilon^{(\frac{1-\xi}{\xi})}$.
Collecting the above results, one finds
\begin{equation}
\label{scaling-H}
h^{st}(\epsilon) \sim
 {\#(\epsilon) \over \tau_N} \sim \epsilon^{-\frac{d-\xi+1}{\xi}}\;.
\end{equation}
For the physical case $d=3$, $\xi=1/3$, one obtains
\begin{equation}
h^{st}(\epsilon) \sim \epsilon^{-11}\;.
\label{32}
\end{equation}
The above result, has already been  obtained in \cite{GW93} with a
different consideration. By denoting with $v_k$ the typical 
velocity at the Kolmogorov scale $\eta$, one has that 
Eq. (\ref{32})  holds in the inertial range, i.e.,
$\epsilon \ge v_k \sim Re^{-1/4}$, while for $\epsilon \le v_k$,
$h^{st}(\epsilon)=$ constant $\sim Re^{11/4}$. 
Let us now discuss the physical implications of (\ref{scaling-H}).

Consider an
alternative way to compute the $\epsilon$-entropy of the field
$v({\bf x},t)$:
divide the $d$-volume in  boxes of edge length $\ell(\epsilon)\sim
\epsilon^{1/\xi}$ and look at the signals $v({\bf x}_\alpha,t)$,
where the ${\bf x}_\alpha$ are the centres of the boxes.
In each  ${\bf x}_\alpha$, we have a time record whose
$\epsilon$-entropy is
\begin{equation}
h^{(\alpha)}(\epsilon) \sim \epsilon^{-1/\xi}\;
\label{entrotime}
\end{equation}
because of the scaling (\ref{scaling-t}). In (\ref{entrotime})
we use the symbol
$h^{(\alpha)}$ to denote the entropy of the temporal evolution
of the velocity field measured in ${\bf x_{\alpha}}$. 
Therefore, $h^{st}(\eps)$ will be obtained summing up
all the "independent" contributions (\ref{entrotime}), i.e.
\begin{equation}
h^{st}(\epsilon) \sim {\cal N}(\epsilon) h^{(\alpha)}(\epsilon)
 \sim {\cal N}(\epsilon) \epsilon^{-1/\xi}\;,
\end{equation}
where ${\cal N}(\epsilon)$ is the number of the independent cells.
It is easy to understand that the simplest assumption ${\cal N}(\epsilon)\sim
l(\epsilon)^d\sim \epsilon^{d/\xi}$ gives a wrong result, indeed one obtains
\begin{equation}
h^{st}(\epsilon) \sim \epsilon^{-\frac{d+1}{\xi}}\;,
\end{equation}
which is not in agreement with (\ref{scaling-H}).
In order to obtain the correct result (\ref{32}) it is necessary to assume
\begin{equation}
{\cal N}(\epsilon)\sim l(\epsilon)^{{\cal D}}\;,
\end{equation}
with ${\cal D}=d-\xi$.
In other words, one has to consider that the sweeping implies a 
nontrivial spatial correlation,
quantitatively measured by the exponent ${\cal D}$, which can be
considered as a sort of ``entropy'' dimension.
Incidentally, we note that ${\cal D}$ has the same numerical 
value as the fractal dimensions of the iso-surfaces $v=const.$ \cite{man}.
From this observation, at first glance, one could conclude  that the above result is
somehow trivial since it is simply related to a geometrical fact.
However, a closer inspection reveals that this is not true.
Indeed, one can construct a 
self-affine field with spatial scaling $\xi$ and 
thus with the fractal dimension of the iso-surfaces $v=const.$ given by
 $d-\xi$ for geometrical reasons, while ${\cal D}=d$.
Such a process can be simply obtained by eliminating the sweeping, i.e.,
\begin{equation}
v({\bf x},t) = \sum_{n=1}^M\sum_{k=1}^{2^{d(n-1)}} a_{n,k}(t)\,
 \varphi \! \left({{\bf x}-{\bf x}_{n,k} \over \ell_n}\right)\,\,,
\label{lagrangian_field}
\end{equation}
where now the ${\bf x}_{n,k}$ are fixed and no longer time-dependent,
while $a_{n,k} \sim (\ell_n)^{\xi}$ but $\tau_n \sim \ell_n$.
For a field described by (\ref{lagrangian_field}) one has that 
 (\ref{scaling-x}) and (\ref{scaling-t}) hold but 
$h^{st}(\epsilon) \sim \epsilon^{-\frac{d+1}{\xi}}$ and ${\cal D}=d$,
while the fractal dimension of the iso-surfaces $v=const.$ is $d-\xi$.
\\
We conclude by noting that it is possible to obtain (see \cite{GW93})
the scaling~(\ref{scaling-H}) using equation~(\ref{lagrangian_field}),
i.e. ignoring the sweeping, assuming $\tau_n \sim (\ell_n)^{1-\xi}$
and $a_{n,k} \sim (\ell_n)^\xi$, this corresponds to take separately the
proper temporal and spatial spectra.
However, this is not completely satisfactory
since one has not the proper scaling in one fixed point,
(see eq.~(\ref{entrotime}) the only way to obtain this is 
through the sweeping).

\section{Conclusion} 
\label{Sec:6} 
In this paper we have discussed a method, based on the analysis of the
exit-time statistics, for the computation of the $\epsilon$-entropy.
The basic idea is to look at a sequence of data 
not at fixed sampling time but only when the fluctuation in the signal 
is larger than some fixed threshold, $\epsilon$. 
This procedure allows a remarkable improvement of the
possibility to compute $(\epsilon, \tau)$-entropy,
which is well represented by the exact results~(\ref{epsent})
and the bounds~(\ref{bound-entro}).

This approach is particularly suitable in all the systems
without a unique characteristic time. In these cases 
the method based on a coarse-grained dynamics on a
fixed $\et$ grid does not work very efficiently since words of very
long size are involved.

On the basis of the coding in terms of the exit-time events we are
able to give significant lower and upper bounds to the
$\epsilon$-entropy.

We have applied the method to different systems: chaotic diffusive
maps, intermittent maps showing sporadic chaos, self-affine and
multi-affine stochastic processes, and experimental turbulent data.

Applying the multifractal formalism one predicts the scaling
$h(\epsilon)\sim \epsilon^{-3}$ for time measurement of velocity in
one point in turbulent flows. This power law does not depend on the
intermittent corrections and has been confirmed by the experimental
data analysis results.

Moreover we have shown the connection of the Taylor-frozen
hypothesis and the $\eps$-entropy: the sweeping implies a nontrivial
spatial correlation, quantitatively measured by an ``entropy''
dimension ${\cal D}=8/3$.

\section{Acknowledments}
We acknowledge useful discussions with G.~Boffetta and A.~Celani and
the encouragement by B.~Marani. We are deeply indebted to Y. Gagne and to
G. Ruiz-Chavarria for having provided us with their experimental data. 
This work has been partially supported by INFM (PRA-TURBO) and by the
European Network {\it Intermittency in Turbulent Systems} (contract
number FMRX-CT98-0175) and the MURST {\it cofinanziamento 1999} 
``Fisica statistica e teoria della materia condensata''. 
M.A. is supported by the European Network {\it 
Intermittency in Turbulent Systems}.

\section{Appendix A}

In this Appendix we recall some recently obtained results on the
generation of multi-affine stochastic signals~\cite{bbcpvv93,bbccv98}.
The goal is to have a stochastic process whose scaling properties are
fully under control. The first step consists in generating a
$1$-dimensional signal and the second in decorating it such as to
build the most general $(d+1)$-dimensional process, $v({\bf x},t)$,
with given scaling properties in time and in space. 
\\ 
As for the simplest case of a $1$-dimensional system there are at
least two different kind of algorithms. One is based on a dyadic
decomposition of the signal in a wavelet basis with a suitable
assigned series of stochastic coefficients \cite{bbcpvv93}. The second
is based on a multiplication of sequential Langevin-processes with a
hierarchy of different characteristic times \cite{bbccv98}. 
\\
The first procedure suits particularly appealing for the modelisation
of spatial turbulent fluctuations, because of the natural identification 
between wavelets and eddies in the physical space. The second one, on
the other hand, looks more appropriate for mimicking the turbulent
time evolution in a fixed point of the space, because of its sequential nature.
\\
Let us first summarise the main ingredient of both and then briefly 
explain how to merge them in order to have a realistic
spatial-temporal multi-affine signal. \\
A non-sequential algorithm for $1$-dimensional 
multi-affine signal in $[0,1]$, $v(x)$,  can be defined as~\cite{bbcpvv93}:
\begin{equation}
v(x) = \sum_{n=1}^N\sum_{k=1}^{2^{(n-1)}} a_{n,k}\,
                   \varphi\!\left(\frac{x-x_{n,k}}{\ell_n}\right)
\label{diadic1}
\end{equation}
where we have introduced a set of reference scales $\ell_n=2^{-n}$
and the function $\varphi(x)$ is a wavelet-like function  \cite{wavelets},
i.e. of zero mean and rapidly decaying in both real space and Fourier-space. 
The signal $v(x)$ is built in terms of a superposition of fluctuations,
$\varphi((x-x_{n,k})/\ell_n)$ of characteristic width $\ell_n$ and centred 
in different points of $[0,1]$,  $x_{n,k} = (2k+1)/2^{n+1}$.
In \cite{bbccv98} it has been proved that provided the coefficients
$a_{n,k}$ are chosen by a random multiplicative process, i.e. 
the daughter is given in terms of the mother by a random
process, $a_{n+1,k'} = X a_{n,k}$ with $X$ a random number i.i.d.
for any $\{n,k\}$, then the result
of the superposition  is a multi-affine function with given 
scaling exponents, namely:
$$
\langle \langle |v(x+R)-v(x)|^p \rangle\rangle \sim R^{\,\zeta(p)}\,\,,
$$ 
with $\zeta(p) = -p/2 - \log_2 \langle X^p \rangle$ and $ \ell_N \leq R \leq 1$.
In this Appendix $\langle \cdot \rangle$ indicates the average over the probability
distribution of the multiplicative process. Besides the rigorous proof,
 the rationale for the previous result is simply that 
due to the hierarchical organisation of the fluctuations one may  easily
estimate that the term dominating the 
expression of a velocity fluctuation at scale $R$, in
(\ref{diadic1}) is given by the couple of indices $\{n,k\}$ 
such that  $n \sim log_2(R)$ and $x \sim x_{n,k}$, i.e. 
$v(x+R)-v(x) \sim a_{n,k}$. 
The generalisation (\ref{diadic1}) to d-dimensional fields is given by:
$$
v({\bf x}) = \sum_{n=1}^N\sum_{k=1}^{2^{d(n-1)}} a_{n,k}\,
 \varphi\!\left(\frac{{\bf x}-{\bf x}_{n,k}}{\ell_n}\right)\,\,,
$$
where now the coefficient $a_{n,k}$ are given in terms of a
d-dimensional dyadic multiplicative process. This class of stochastic
fields has been of great help in mimicking simultaneous spatial
fluctuations of turbulent flows. 
\\ 
On the other hand, as previously said, sequential algorithms look more
suitable for mimicking temporal fluctuations. Let us now discuss how
to construct these stochastic multi-affine fields. With the
application to time-fluctuations in mind, we will denote now the
stochastic 1-dimensional functions with $u(t)$. The signal $u(t)$ is
obtained by a superposition of functions with different characteristic
times, representing eddies of various sizes~\cite{bbccv98}:
\begin{equation}
u(t)=\sum_{n=1}^N u_n(t) \; .
\label{eq:decomp}
\end{equation}
The functions $u_n(t)$ are defined by the multiplicative process
\begin{equation}
u_n(t)=g_n(t)x_1(t)x_2(t)\ldots x_n(t) \; ,
\label{def:mult}
\end{equation}
where the $g_n(t)$ are independent stationary random processes, 
whose correlation times are supposed to be $\tau_n=(\ell_n)^\alpha$, 
where $\alpha = 1-\xi$ (i.e. $\tau_n$ are the eddy-turn-over time at scale $\ell_n$) 
in the quasi-Lagrangian reference frame~\cite{lag2} and $\alpha = 1$ if one considers
$u(t)$ as the time signal in a given point, 
and $\langle g_n^2 \rangle = (\ell_n)^{2\xi}$,
where $\xi$ is the H\"older exponent. For a signal mimicking 
a turbulent flow, ignoring
intermittency, we would have  $\xi=1/3$. 
Scaling will appear for all time delays larger than the UV 
cutoff $\tau_N$ and smaller than the IR cutoff $\tau_1$.
The $x_j(t)$ are independent, positive defined, identical distributed
random processes whose time correlation decays with the characteristic
time $\tau_j$. The probability distribution of $x_j$ determines the
intermittency of the process.

The origin of (\ref{def:mult}) is fairly clear in the context of fully
developed turbulence. Indeed we can identify $u_n$ with the velocity 
difference at scale $\ell_n$ and $x_j$ with 
$(\varepsilon_j/\varepsilon_{j-1})^{1/3}$, where
$\varepsilon_j$ is the energy dissipation at scale $\ell_j$.

The following arguments show, that the process defined
according to (\ref{eq:decomp},\ref{def:mult}) is multi-affine:
Because of the fast decrease of the correlation times
$\tau_j=(\ell_j)^\alpha$, the characteristic time of $u_n(t)$ is of the
order of the shortest one, i.e., $\tau_n=(\ell_n)^\alpha$.  Therefore, the
leading contribution to the structure function $\tilde{S}_q(\tau) =
\langle\langle|u(t+\tau)-u(t)|^q\rangle\rangle$ with $\tau \sim
\tau_n$ stems from the $n$-th term in (\ref{eq:decomp}).  This can be
understood nothing that in the sum $u(t+\tau)-u(t) = \sum_{k=1}^N
[u_k(t+\tau)-u_k(t)]$ the terms with $k \le n$ are negligible because
$u_k(t+\tau) \simeq u_k(t)$ and the terms with $k \ge n$ are
sub-leading.  Thus one has:
\begin{equation}
\tilde{S}_q(\tau_n) 
\sim \langle |u_n|^q \rangle \sim 
\langle |g_n|^q \rangle \langle x^q \rangle^n
\sim \tau_n^{\frac{\xi q}{\alpha} - \frac{\log_2\langle x^{q} \rangle}{\alpha}} 
\end{equation}
and therefore for the scaling exponents:
\begin{equation}
\zeta_q={\xi q \over \alpha} - {\log_2\langle x^{q} \rangle \over \alpha} \; .
\label{eq:zq}
\end{equation}
The limit of an affine function can be obtained when 
all the $x_j$ are equal to $1$. A proper proof of these result
can be found in \cite{bbccv98}. 
\\
Let us notice at this stage that the previous ``temporal'' signal for
$\alpha = 1 - \xi$ is a good candidate for a velocity measurements in
a Lagrangian, co-moving, reference frame (see body of the article).
Indeed, in such a reference frame the temporal decorrelation
properties at scale $\ell_n$ are given by the eddy-turn-over times
$\tau_n=(\ell_n)^{1-\xi}$.  On the other hand, in the laboratory
reference frame the sweeping dominates the time evolution in a fixed
point of the space and we must use as characteristic times of the
processes $x_n(t)$ the sweeping times $\tau_n^{(s)} = \ell_n$, i.e.,
$\alpha=1$.

\section{Appendix B}
We have now all the ingredients to perform a merging of temporal and
spatial properties of a turbulent signal in order to define
 stochastic processes able to reproduce in a realistic way both
spatial and temporal fluctuations in a Lagrangian reference frame. 
We just have to merge in a proper way the two previous algorithms. \\
For example, for a d-dimensional multi-affine field such as, say, 
one of the three components of a turbulent field
in a Lagrangian reference frame we can use the following model:
\begin{equation}
v_L({\bf x},t) = \sum_{n=1}^N\sum_{k=1}^{2^{d(n-1)}} a_{n,k}(t)\,
 \varphi \! \left(\frac{{\bf x}-{\bf x}_{n,k}}{\ell_n}\right).
\label{lagrangian_field1}
\end{equation}
where the temporal dependency of $ a_{n,k}(t)$ is chosen
following the sequential algorithm while its spatial part are given
by the dyadic structure of the non-sequential algorithm.
 In (\ref{lagrangian_field1})
we have used the notation $v_L({\bf x},t)$ in order to stress the typical
Lagrangian character of such a field.  \\
We are now also able to guess a good candidate for the same field
measured in the laboratory-reference frame, i.e. where the time
properties are dominated by the sweeping of small scales by large scales. 
Indeed, it is enough to physically reproduce the sweeping effects  by
allowing the centre of the wavelets-like functions used to mimic the
eddies-like turbulent structures to move according a swept-dynamics.

To do so, let us define the Eulerian model:
\begin{equation}
\label{eulerian_field}
v_E({\bf x},t) = \sum_{n=1}^N\sum_{k=1}^{2^{d(n-1)}} a_{n,k}(t)
 \varphi\left(\frac{{\bf x}-{\bf x}_{n,k}(t)}{\ell_n}\right).
\end{equation}
where the difference with the previous definition is in the
temporal dependency of the centres of the wavelets, ${\bf x}_{n,k}(t)$.
According to the Richardson-Kolmogorov cascade picture, one assumes
that sweeping is present, i.e.,
${\bf x}_{n,k}={\bf x}_{n-1,k^\prime}+{\bf r}_{n,k}$ where
$(n,k^\prime)$ labels the "mother" of the
$(n,k)$-eddy and ${\bf r}_{n,k}$
is a stochastic vector which depends
on ${\bf r}_{n-1,k^\prime}$  and evolves with characteristic time
$\tau_n \propto (\ell_n)^{1-\xi} $.
 Furthermore, its norm is $O(\ell_n)$:
$c_1 < |{\bf r}_{n,k}|/\ell_n < c_2$ where $c_1$ and $c_2$ are constants of
order one. \\
We now see that if we measure in one fixed spatial point a 
fluctuations over a time delay $\delta t$, is like to measure
a simultaneous fluctuations at scale separation $R=U_0\delta t$, 
i.e. due to the sweeping
the main contribution to the sum will be given by 
the terms with scale-index $n = \log_2(R=U_0\delta t)$ while the 
temporal dependency  of the coefficients $a_{n,k}(t)$
will be practically frozen on that time scale. This happens because
in presence of the sweeping the main contribution is given by
the displacement of the centre at large scale, i.e. 
$\delta r_0 = |{\bf r_0}(t+\delta t) - {\bf r_0}(t)|\sim U_0 \delta t$,
and the eddy turnover time at scale $\ell_n$
is $O((\ell_n)^{1-\xi})$ always large that the sweeping
time $O(\ell_n)$ at the same scale.
\\
In the previous discussion for sake of simplicity we did not consider
the incompressibility condition.  However one can take into account
this constraint by the projection on the solenoidal space.
\\
In conclusion we  have a way to build up a synthetic signal
with the proper Eulerian (laboratory) properties, i.e. with sweeping,
and also with the proper Lagrangian properties. 


\newpage

\begin{figure}[htb]
\centerline{\epsfig{figure=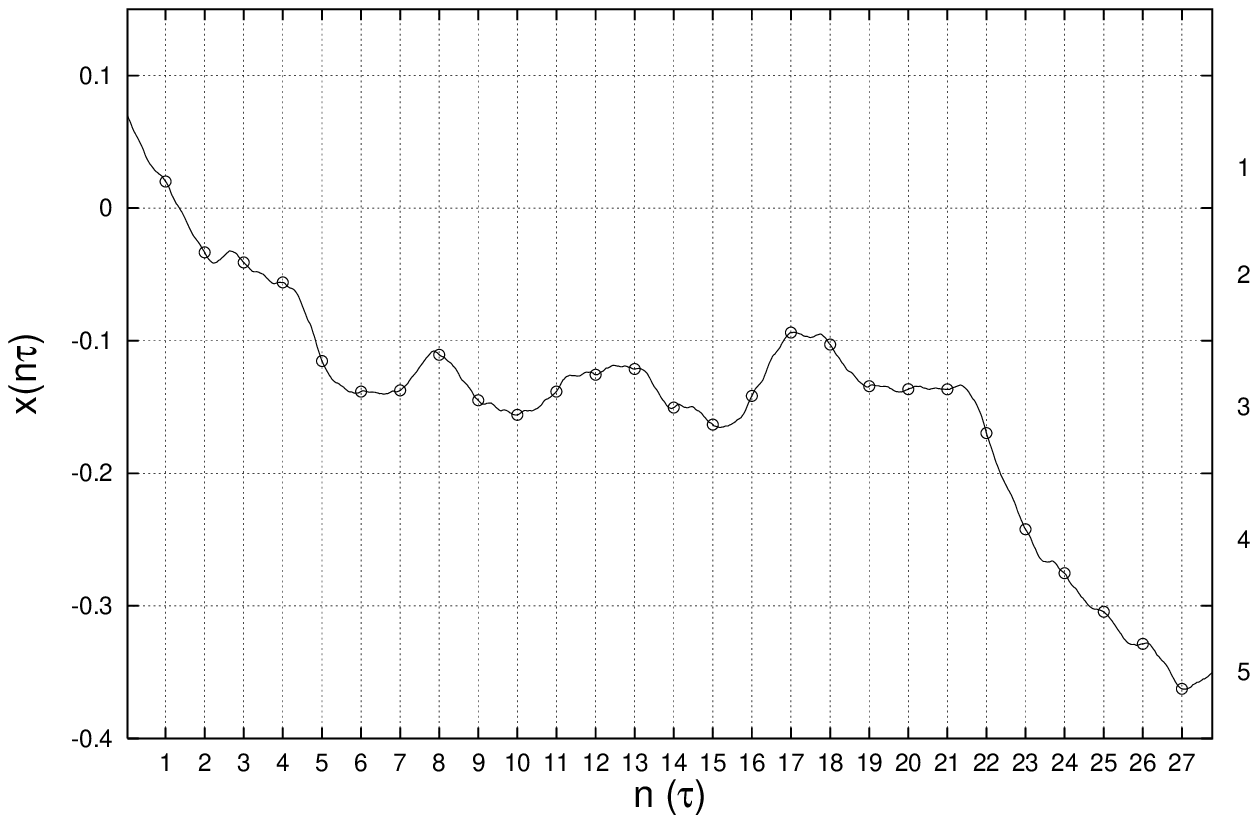,width=0.55\textwidth,angle=0}}

\caption{Sketch of the coding procedure described in Section II.
On the given $\et$-grid the symbolic sequence is 
$W_0^{27}\et=(1,2,2,2,3,3,3,3,3,3,3,3,3,3,3,3,2,3,3,3,3,3,4,4,5,5,5)$.}
\label{fig:coding_naive}
\end{figure}

\begin{figure}[hbt]
\centerline{\epsfig{figure=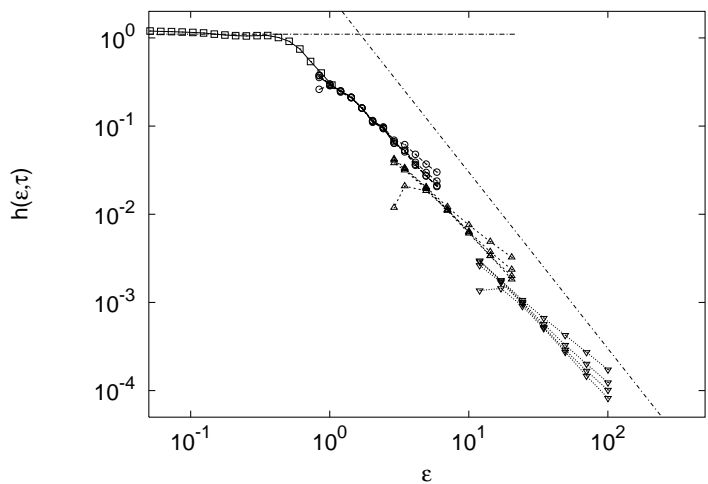,width=0.6\textwidth,angle=0}}

\caption{ Numerically evaluated $(\epsilon,\tau)$-entropy
for the map (\ref{eq:mappa}) with $p=0.8$ computed with the
Grassberger-Procaccia algorithm $[6]$ at $\tau=1$ ($\circ$), 
$\tau=10$ ($\bigtriangleup$) and $\tau=100$ ($\bigtriangledown$) 
and different block length ($n=4,8,12,20$).
The boxes ($\Box$) give the entropy computed with $\tau=1$ by
using periodic boundary condition over $40$ cells. The latter is
necessary in order to compute the Lyapunov exponent
$\lambda=h_{KS}=1.15$.
The straight lines correspond to the two
asymptotic behaviours, $h(\epsilon)=h_{KS}$ and $h(\epsilon) \sim
\epsilon^{-2}$. }
\label{fig:diffmapusual}
\end{figure}

\begin{figure}[hbt]
\centerline{\epsfig{figure=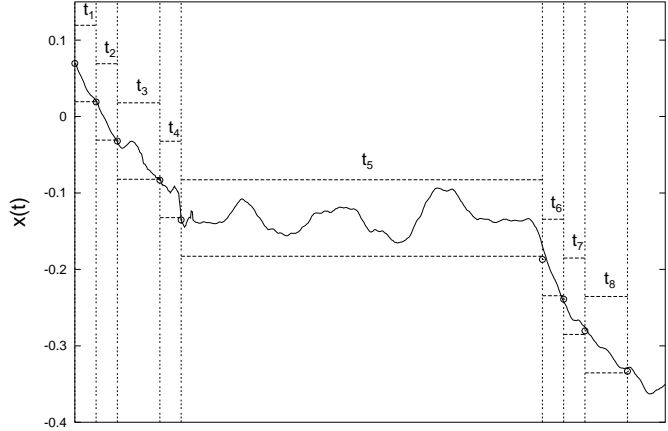, width=0.55\textwidth,angle=0}}

\caption{The same signal as in  Fig.~1,  with the exit-time coding
of the same precision $\epsilon$.  The symbolic sequence obtained with
the exit time method is $\Omega_0^{27}=[(t_1,-1);(t_2,-1);(t_3,-1);
(t_4,-1);(t_5,-1);(t_6,-1);(t_7,-1);(t_8,-1)]$.}
\label{fig:coding_exit}
\end{figure}

\begin{figure}
\label{fig:diffmapet}
\centerline{\epsfig{figure=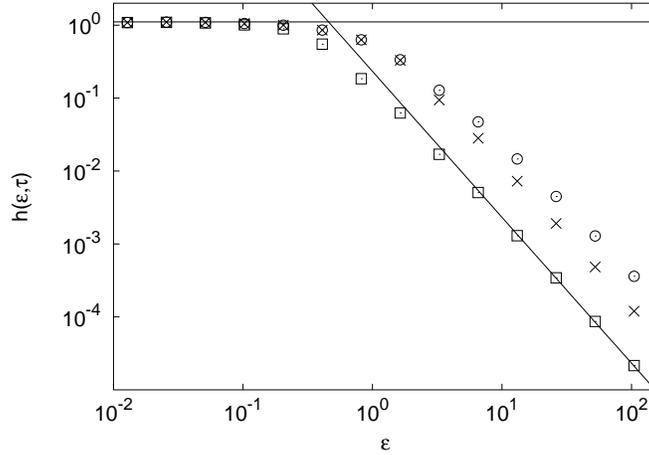, width=0.55\textwidth,angle=0}}

\caption{Numerically computed lower bound ($\Box$) and
upper bound (with $\tau=1$) ($\circ$)
of $h(\epsilon)$ according to Eq.~(\ref{bound-entro}),
for the map (\ref{eq:mappa}) with the same parameters
as in Fig.~\ref{fig:diffmapusual}.
The two straight lines correspond
 to the asymptotic behaviours as in Fig.~\ref{fig:diffmapusual}.
The crosses (x) mark the values of  the $(\epsilon,\tau)$-entropy
$h^\Omega(\epsilon,\tau)/\langle t(\epsilon)\rangle$
with $\tau=0.1\, \langle t(\epsilon)\rangle$.}
\end{figure}

\begin{figure}
\centerline{\epsfig{figure=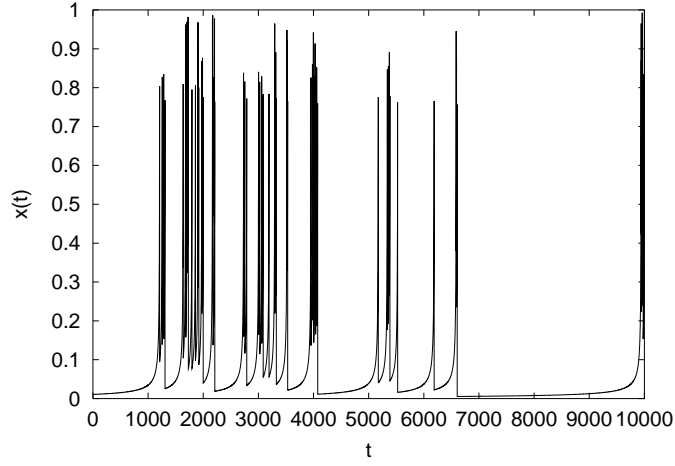,width=0.55\textwidth,angle=0}}

\protect\caption{Typical evolution of the intermittent map (17) for $z=2.5$ and $a=0.5$.}
\label{fig:traj}
\end{figure}

\begin{figure}
\epsfig{figure=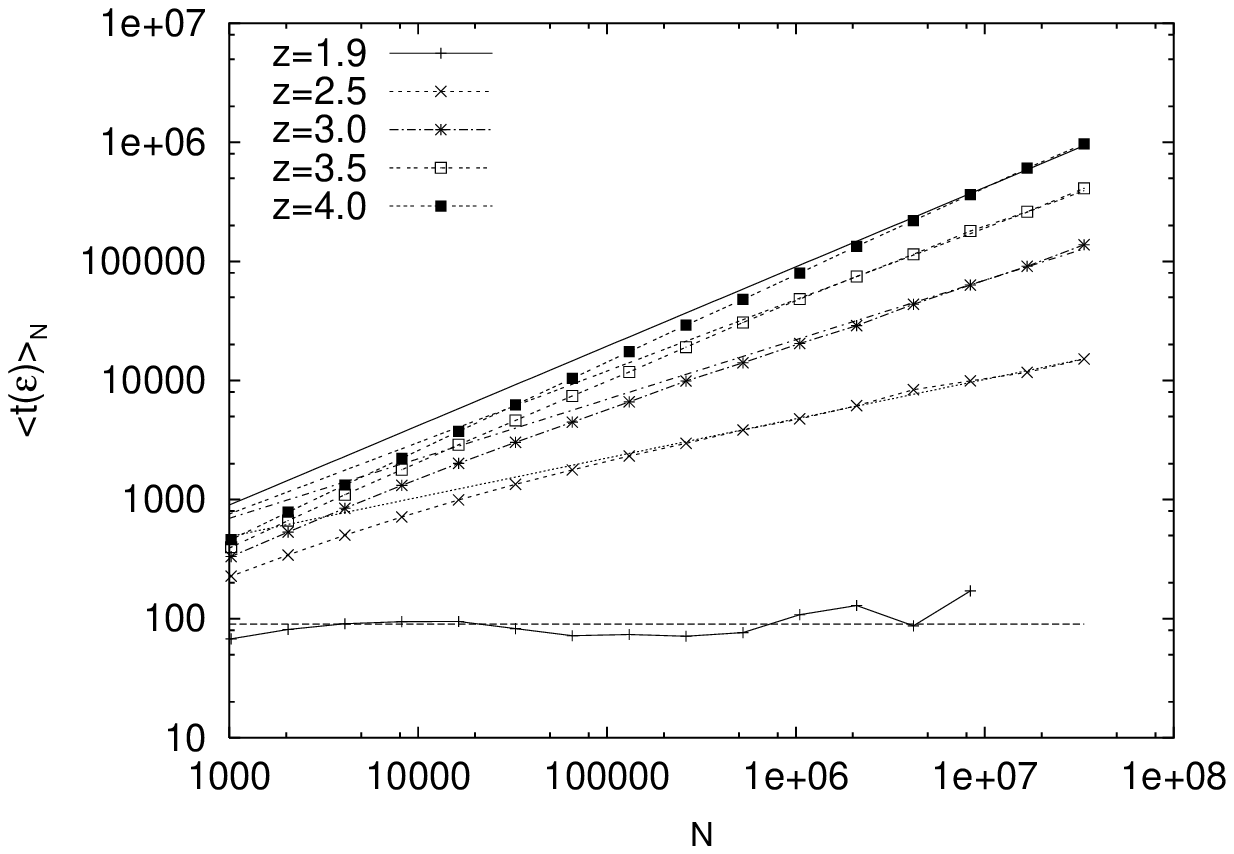,width=0.45\textwidth, height=0.26\textheight}
\epsfig{figure=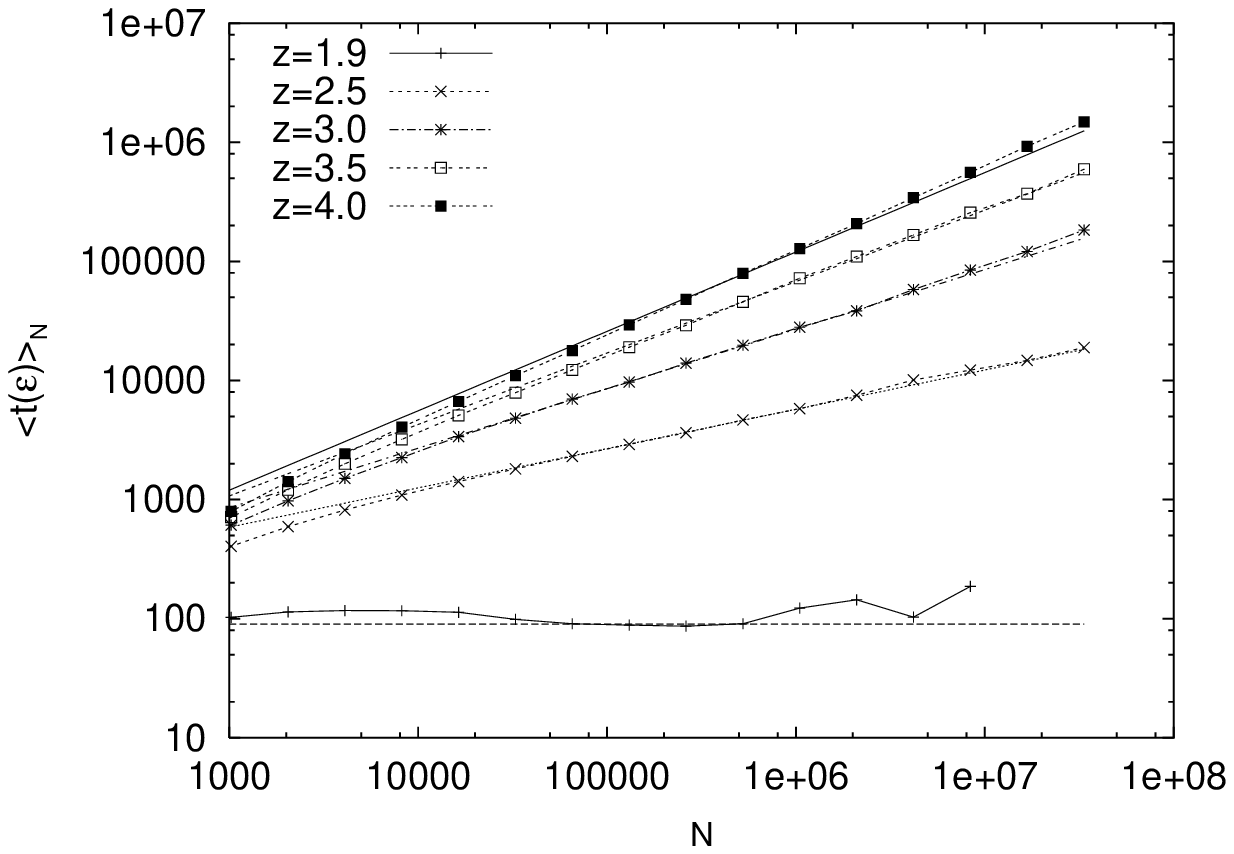,width=0.45\textwidth, height=0.26\textheight}

\protect\caption{$\langle t(\epsilon)\rangle_N$ versus $N$ for the intermittent map
(\protect\ref{eq:41.1}) at $\epsilon=0.001$ (left) and $\epsilon=0.243$ (right) 
for different $z$  and $a=0.5$. The straight lines indicate
the power law (\ref{eq:tave}). The average $\langle
t(\epsilon)\rangle_N$ has been obtained by averaging over $10^4$ different
trajectories of length $N$, this average is necessary because of
the poor statistics caused by the singularity near the origin.
For $z < 2$, $\langle t(\epsilon)\rangle_N$ does not depend on $N$,
$\rho(x)$ is normalisable, the motion is chaotic and $H_N / N$ is constant.}
\label{fig:intermap}
\end{figure}
\newpage
\begin{figure}
\epsfig{figure=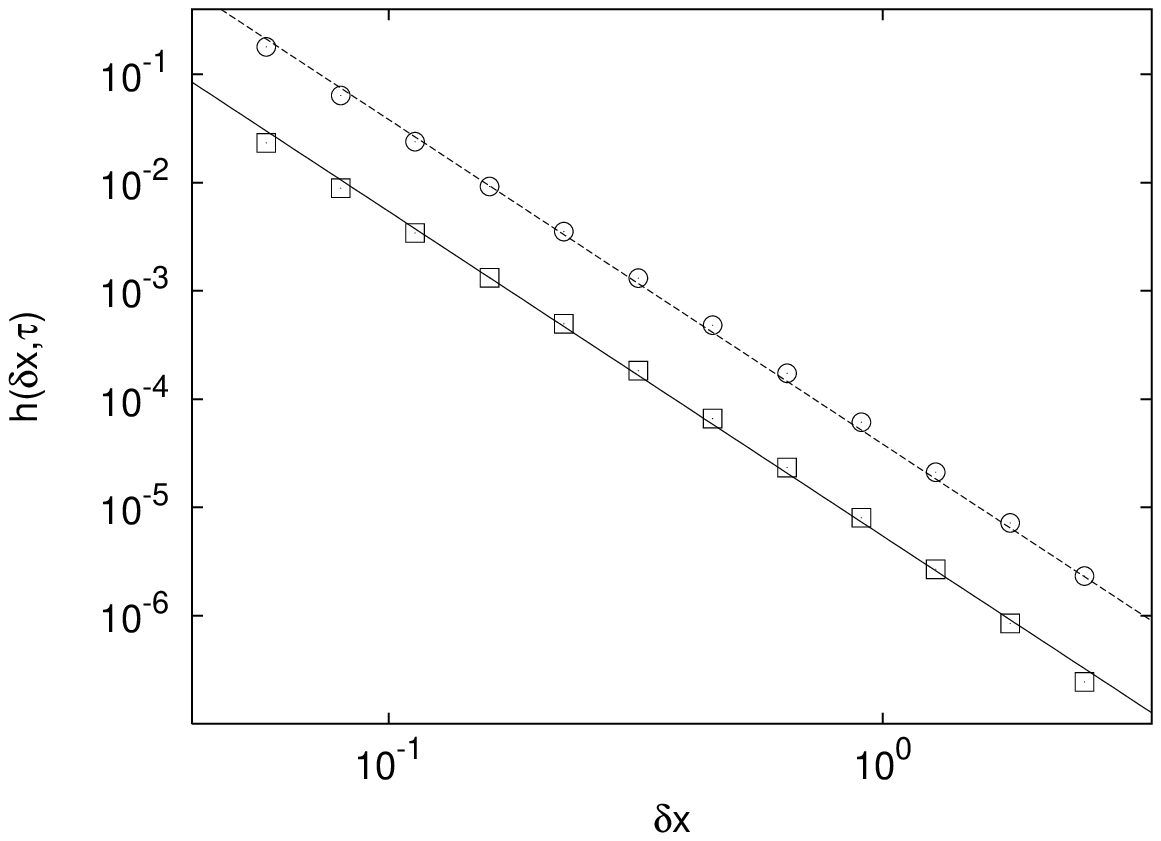,width=0.45\textwidth, height=0.26\textheight}
\epsfig{figure=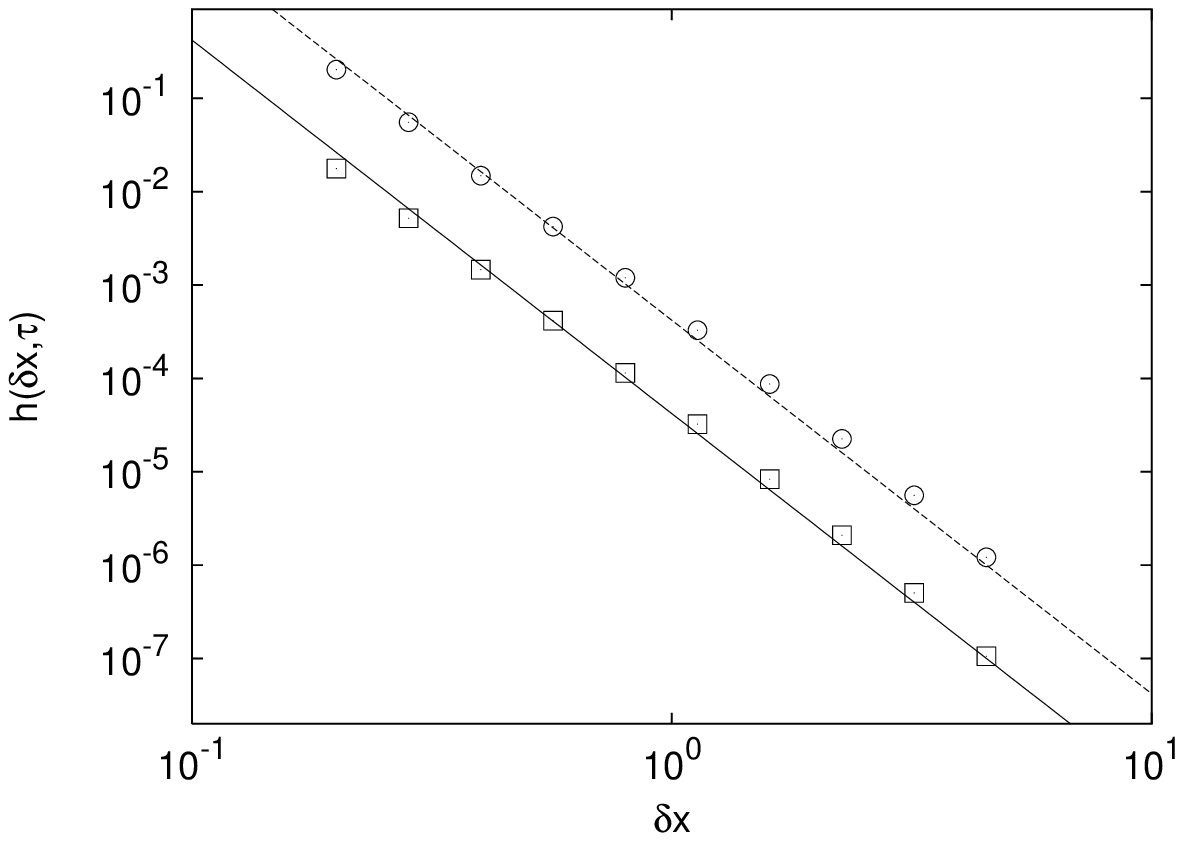,width=0.45\textwidth, height=0.26\textheight}

\protect\caption{Numerically computed lower ($\Box$) and upper bound 
($\circ$) for the $(\epsilon,\tau)$-entropy in the case
of a self-affine signal with $\xi=1/3$ (left) and $\xi=1/4$ (right),
with $\tau=0.1\langle t(\epsilon)\rangle$.  
The two straight lines show the scaling $\epsilon^{-3}$ and $\epsilon^{-4}$
for the left and the right figure, respectively.}
\label{figaffine}
\end{figure}

\begin{figure}
\centerline{\epsfig{figure=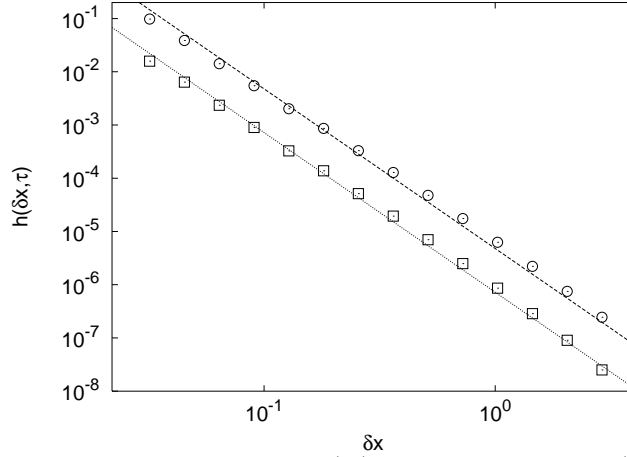,width=0.55\textwidth}}

\protect\caption{Numerically computed lower bound ($\Box$) 
and upper bound ($\circ$), with $\tau=0.1\langle t(\epsilon)\rangle$ 
for the $(\epsilon,\tau)$-entropy in the case
of a multiaffine signal with $\zeta(3)=1$.
The signal has been obtained with the method of Ref.~[35]
(see also Appendix A) using a $D(\xi)$ which fits experimental data
at large Reynolds number.
The two straight lines show the theoretical scaling $\epsilon^{-3}$.}
\label{figmulti}
\end{figure}
\newpage
\begin{figure}
\epsfig{figure=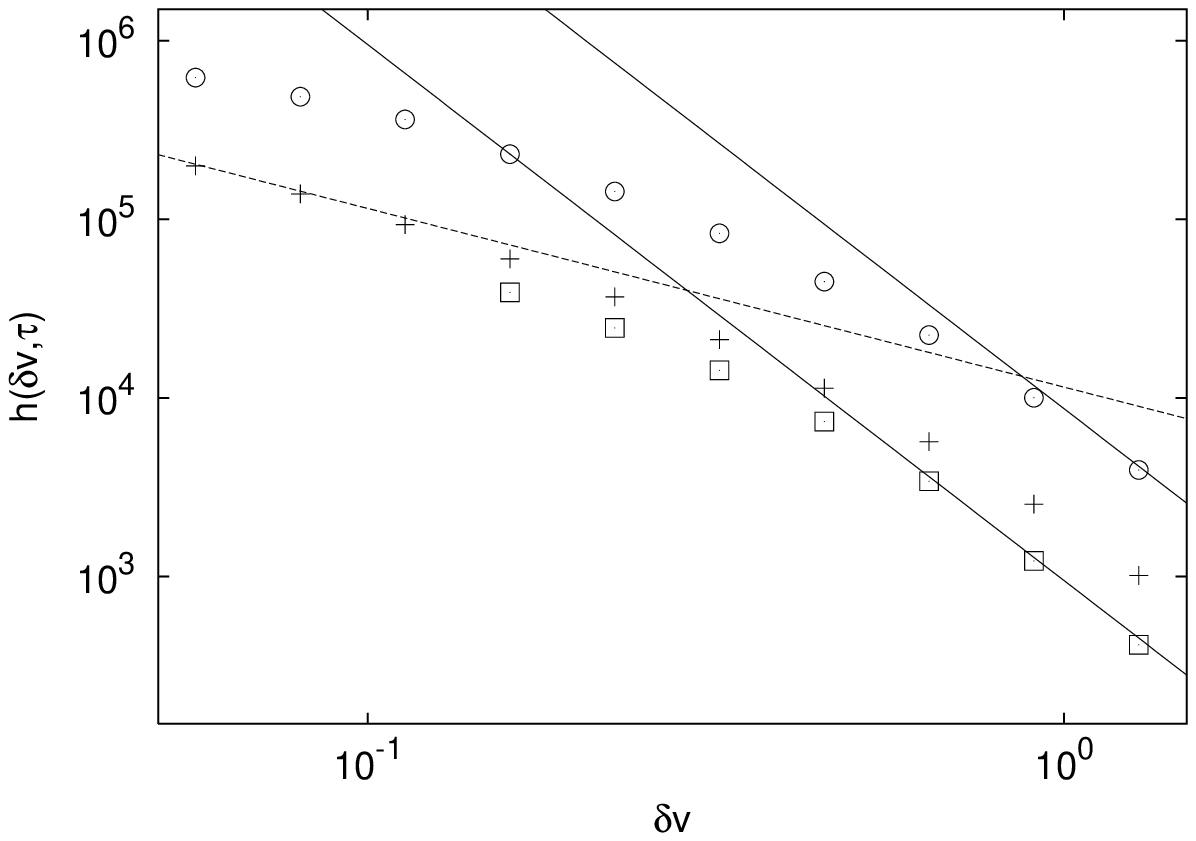,width=0.45\textwidth, height=0.26\textheight}
\epsfig{figure=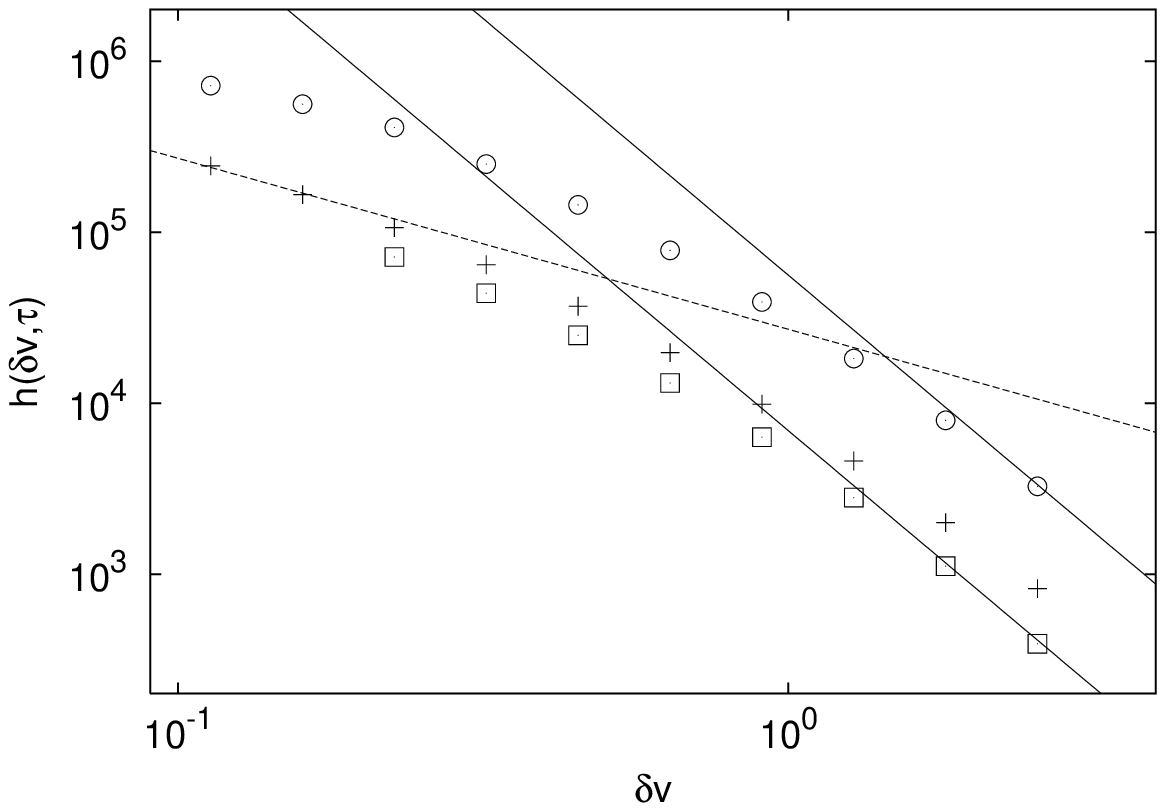,width=0.45\textwidth, height=0.26\textheight}

\protect\caption{Numerically computed lower bound ($\Box$) and upper bound 
($\circ$), with $\tau=0.1\langle t(\epsilon)\rangle$ 
for the $(\epsilon,\tau)$-entropy in the case
of Lyon turbulent data (left) and Modane turbulent data (right).
We also show $\langle t(\delta v) \rangle^{-1}$ (+)
and its trivial dissipative scaling $\delta v^{-1}$ (dashed line).
The full line follows the scaling $\delta v^{-3}$
for the $\eps$-entropy, as predicted in Eq.~(\ref{predturb}).}
\label{fig:turbo}
\end{figure}

\begin{figure}[thb]
\label{fig:giostra}
\centerline{\epsfig{figure=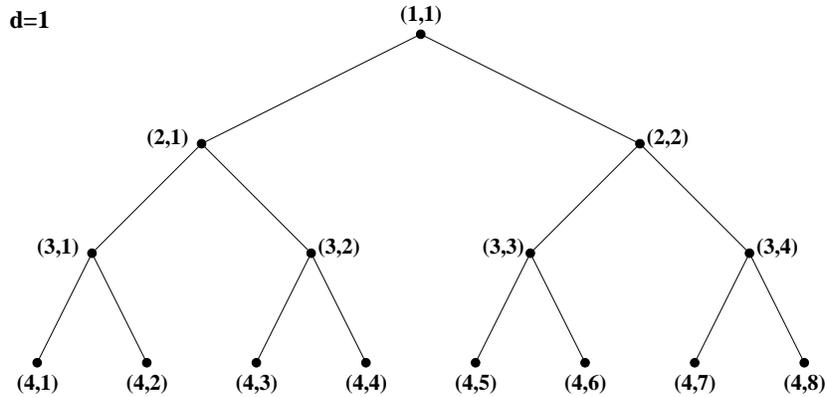,width=0.7\textwidth,angle=0}}

\protect\caption{Branching process for the multiplicative model (we only show
the $d=1$ case), as described in the main text.}
\end{figure}

\begin{figure}[thb]
\label{fig:tree}
\centerline{\epsfig{figure=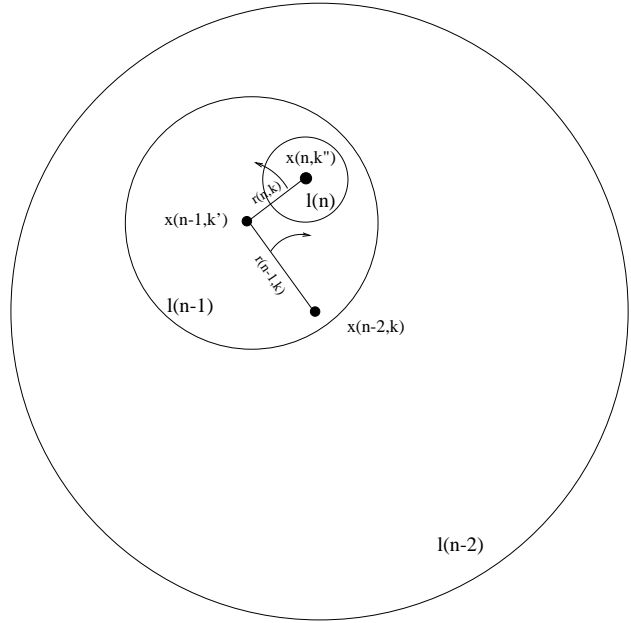,width=0.5\textwidth,angle=0}}

\vspace{0.1truecm}

\protect\caption{Sketch of the construction of the synthetic turbulent field.
Circles represent symbolically the eddies on the scale $n$, $n-1$, $n-2$.
The centers of the eddies are denoted by $x$,  $r$ indicates the distances
between subsequent generations and the arrows hint to the sweeping motion.}
\end{figure}

\end{document}